\documentclass[journal=jctcce,manuscript=article]{achemso}
\usepackage[version=4]{mhchem} 
\usepackage{graphicx}
\usepackage{mciteplus}
\usepackage{booktabs}
\usepackage{multirow}
\usepackage{hyperref}
\hypersetup{colorlinks=true,linkcolor=blue,citecolor=blue,urlcolor=blue}
\usepackage{soul}
\usepackage[normalem]{ulem}
\usepackage[dvipsnames]{xcolor}

\def\QE{\textsc{Quantum ESPRESSO}\,}

\author{Eric Macke}
\affiliation{Faculty of Production Engineering, Bremen Center for Computational Materials Science and MAPEX Center for Materials and Processes, Hybrid Materials Interfaces Group, University of Bremen, Am Fallturm 1, 28359 Bremen, Germany}
\email{emacke@uni-bremen.de}

\author{Iurii Timrov}
\affiliation{Theory and Simulation of Materials (THEOS) and National Centre for Computational Design and Discovery of Novel Materials (MARVEL), \'Ecole Polytechnique F\'ed\'erale de Lausanne, CH-1015 Lausanne, Switzerland}
\altaffiliation{Present address: Laboratory for Materials Simulations (LMS), Paul Scherrer Institut (PSI), CH-5232 Villigen PSI, Switzerland}

\author{Nicola Marzari}
\affiliation{Theory and Simulation of Materials (THEOS) and National Centre for Computational Design and Discovery of Novel Materials (MARVEL), \'Ecole Polytechnique F\'ed\'erale de Lausanne, CH-1015 Lausanne, Switzerland}
\alsoaffiliation{University of Bremen Excellence Chair, Bremen Center for Computational Materials Science}

\author{Lucio Colombi Ciacchi}
\affiliation{Faculty of Production Engineering, Bremen Center for Computational Materials Science and MAPEX Center for Materials and Processes, Hybrid Materials Interfaces Group, University of Bremen, Am Fallturm 1, 28359 Bremen, Germany}

\title[Orbital-resolved DFT\textit{\textbf{+U}} for molecules and solids]{Orbital-resolved DFT\textit{\textbf{+U}} for molecules and solids}

\abbreviations{DFT,DFT+U,DFT+U+V,DFT+U+J,cDFT,LR-cDFT,LS,HS,FCE,SIE,PWL,KS,MO,TM,NAO,OAO,PAW,MLWF,DFPT,VBM,CBM,CASPT2/CC,AFM,KI,LDA,GGA,DMFT}

\begin{document}



\begin{abstract}
    We present an orbital-resolved extension of the Hubbard $U$ correction to density-functional theory (DFT). 
    Compared to the conventional shell-averaged approach, the prediction of energetic, electronic and structural properties is strongly improved, particularly for compounds characterized by both localized and hybridized states in the Hubbard manifold.
    The numerical values of all Hubbard parameters are readily obtained from linear-response calculations.
    The relevance of this more refined approach is showcased by its application to bulk solids pyrite (\ce{FeS2}) and pyrolusite (\ce{\beta-MnO2}), as well as to six Fe(II) molecular complexes.
    Our findings indicate that a careful definition of Hubbard manifolds is indispensable for extending the applicability of DFT+$U$ beyond its current boundaries. The present orbital-resolved scheme aims to provide a computationally undemanding yet accurate tool for electronic structure calculations of charge-transfer insulators, transition-metal (TM) complexes and other compounds displaying significant orbital hybridization.
\end{abstract}

\section{Introduction}
    \label{sec:intro}
    Hubbard corrections are among the most widely used improvements to approximate Kohn-Sham (KS) density-functional theory (DFT) \cite{hohenberg_inhomogeneous_1964, kohn_self-consistent_1965}.
    The combination of DFT with the Hubbard model~\cite{hubbard_electron_1963}, referred to as DFT+$U$, was inspired by this widely studied model of electron correlations and introduced to improve the description provided by local or semi-local exchange-correlation (xc) functionals, such as the local-density approximation (LDA) and the generalized-gradient approximation (GGA), for the case of Mott-Hubbard insulators~\cite{anisimov:1991, anisimov_density-functional_1993,liechtenstein_density-functional_1995} or more broadly strongly correlated electrons.
    
    It was recognized early on~\cite{cococcioniLDAStudySelected2002,cococcioni_linear_2005} that a simplified, rotationally invariant formulation of DFT$+U$~\cite{dudarev_electron-energy-loss_1998} provided a natural connection to the requirement of piecewise linearity of the exact energy functional\cite{Perdew_dft_1982,yang_pwl_2000}: the Hubbard correction of Dudarev effectively removes the non-linear (almost quadratic) behavior of the total energy with respect to the occupation of the Hubbard manifold, and replaces it with a linear term. In this light, the strength of the $U$ parameter can be determined fully from first principles, and obtained from the second derivative of the energy (once the non-interacting terms are removed) with respect to the occupations~\cite{cococcioniLDAStudySelected2002,cococcioni_linear_2005}, so that the quadratic curvature is removed by DFT$+U$.
    This connection is heuristic, and relies on the (very reasonable) assumption that the localized electrons in the $d$ or $f$ manifold are only weakly interacting with the rest of the electron bath, so that this manifold also follows the condition of piecewise linearity (PWL) that is in principle valid only for the total energy functional\cite{Perdew_dft_1982}. The overstabilization of fractional occupations in standard (semi)-local functionals is driven by the incomplete cancellation from the xc functional to the Hartree term (cancellation that is exact instead in Hartree-Fock), leading to a strong (one-electron) self-interaction component. This is particularly severe for localized electrons, that over-delocalize and over-hybridize with their ligands.
    So, the DFT$+U$ correction should be seen as a self-interaction correction, providing an approximate screened Fock-like contribution; this was stated early on, highlighting how even for molecular systems containing only one site (i.e., one TM atom) the electronic-structure description provided by DFT$+U$ is greatly improved, both qualitatively and quantitatively~\cite{kulikDensityFunctionalTheory2006}. In other words, while the original Hubbard model was concerned with the correlation effects that appear on a lattice, where each site can only contain 0, 1 or 2 electrons, the functional form that the model inspires, once applied to the continuum of the electron gas of approximate DFT, does not any more account for correlations between sites, but counteracts the tendency of strongly localized electrons to hybridize with their ligands, driven by incomplete cancellation in the Hartree term of the one-electron self-interaction.
    This can be easily argued not only by applying DFT$+U$ to molecules~\cite{kulikDensityFunctionalTheory2006}, but also by the simple observation that changing the value of $U$ changes the charge transfer to/from the ligands\cite{zhao_where_2018}, but does not transfer charge to the Hubbard manifolds on the other sites, highlighting the completely different physics of the Hubbard Hamiltonian on a lattice, and the Hubbard $U$ correction in the continuum of Kohn-Sham electrons. Incidentally, strongly correlated materials have almost invariably very localized $d$ or $f$ electrons, so do benefit from DFT$+U$ -- but not because the latter improves the treatment of correlations, but just because it decreases the self-interaction errors (SIE).
    It is important to iterate that PWL is a property that a system must obey on the \textit{global} scale, i.e., with respect to the changes in the total number of electrons\cite{Levy_electron_1982}.
    DFT$+U$, on the other hand, seeks to eliminate the curvature \textit{locally} by restoring the piece-wise linear behavior in a subspace known as the Hubbard manifold~\cite{cococcioni_linear_2005, zhao_global_2016}. The relation between self-interaction and piecewise linearity in many-electron systems is somewhat confusing; in fact, Koopmans functionals~\cite{daboKoopmansConditionDensityfunctional2010,borghiKoopmanscompliantFunctionalsTheir2014} impose (very accurately) piecewise linearity on each orbital in a given system, but do not correct self-interaction; the integer Koopmans functional (KI) provides the same exact total energy of the base functional (LDA or GGA), but a completely different spectrum.
    
    A key challenge in DFT$+U$ arises from the (\textit{a priori}) choice of the value of the on-site Hubbard $U$ parameter.
    Often, $U$ is considered a tunable quantity, adjusted to achieve agreement with experimental results for specific properties of interest such as band gaps, lattice parameters, magnetic moments, or formation enthalpies~\cite{Wang:2006, LeBacq:2004, Aykol:2014, Isaacs:2017}.
    However, this empirical procedure possesses limited predictive capabilities due to its reliance on experimental data; not to mention that certain quantities, like band gaps, are not even meant to be predicted by DFT.
    Alternatively, $U$ values can be computed from first principles. The different methods that have been proposed for this purpose can be classified into three groups: the constrained DFT approach (cDFT)~\cite{dederichs_ground_1984, Mcmahan:1988, Gunnarsson:1989, Hybertsen:1989, Gunnarsson:1990, pickett_reformulation_1998, Solovyev:2005, Nakamura:2006, shishkin_self-consistent_2016}, the constrained random phase approximation (cRPA)~\cite{Springer:1998, Kotani:2000, aryasetiawan_frequency-dependent_2004, Aryasetiawan:2006, karlsson_method_2010}, and Hartree-Fock based approaches~\cite{mosey_ab_2007, mosey_rotationally_2008, Andriotis:2010, Agapito:2015, TancogneDejean:2020, Lee:2020}. 
    The linear-response formulation of constrained DFT (LR-cDFT)~\cite{cococcioni_linear_2005} has become a method of choice for many DFT+$U$ studies~\cite{cococcioni_energetics_2019, Ricca:2019, Sun:2020, kirchner-hall_extensive_2021, Xiong:2021, Zhou:2021, moore_high-throughput_2022, haddadiOnsiteIntersiteHubbard2023, linscott_role_2018}. A recent reformulation of this method in terms of density-functional perturbation theory (DFPT)~\cite{timrov_hubbard_2018, timrov_ultrasoft_2021} has significantly enhanced its success. This reformulation enables the replacement of computationally expensive supercells with primitive unit cells, utilizing monochromatic perturbations. As a result, the computational burden of determining Hubbard parameters is substantially reduced.
      
    As evident from the large variety of available methods, considerable attention has been devoted to the numerical evaluation of $U$ for a given manifold. However, the question how this manifold must be defined, that is, \textit{which} electronic states require on-site Hubbard corrections, is a critical yet often overlooked aspect in DFT$+U$.
    Recalling that the main motivation of Hubbard $U$ corrections lies in the mitigation of local SIE through recovery of PWL of the total energy, the Hubbard manifold should contain those and only those states that substantially contribute to the former.
    Oftentimes, self-interaction occurs in partially occupied $d$ and $f$ shells due to their high electron count and localization; hence, these are the traditional targets of Hubbard $U$ corrections. Nevertheless, self-interaction can also manifest itself in $s$ and $p$ shells, and even in localized molecular orbitals (MO)\cite{leiria_campo_jr_extended_2010}.
    Moreover, the definition of the Hubbard manifold also involves choosing so-called Hubbard-projector functions that suitably represent the states to be corrected (details follow in Section \ref{sec:Hub_projector_functions}).
    In fact, it has been shown that the choice of the projector function can have a stronger impact on the calculation than the numerical value of $U$, especially when orbital overlap is appreciable\cite{mahajan_importance_2021,kirchner-hall_extensive_2021,oregan_projector_2010}. Therefore, it is essential to choose a $U$ value that is consistent with the projectors used (see appendix of Ref. \citenum{kulik_self-consistent_2008}).

    The significance of judiciously selecting the Hubbard manifold becomes particularly evident in the context of optical properties. Although DFT is a total energy theory and KS orbitals have no direct physical meaning except for the highest occupied molecular orbital (HOMO), KS eigenvalues are commonly used as estimates for experimental quasiparticle energies.
    Unfortunately, uncorrected (semi-)local DFT functionals tend to severely underestimate properties such as band gaps and ionization energies; a fact that has been linked to deviations from PWL \cite{daboKoopmansConditionDensityfunctional2010,kraisler_pwl_2013, borghiKoopmanscompliantFunctionalsTheir2014}.
    Thus, Koopmans-compliant functionals\cite{daboKoopmansConditionDensityfunctional2010,borghiKoopmanscompliantFunctionalsTheir2014,linscottKoopmansOpenSourcePackage2023} systematically improve spectral properties by imposing PWL on all orbitals in the system. For example, a recent study by \citeauthor{Nguyen_Koopmans_2018} found an average band gap error of only 0.22\,eV for a test set of 30 semiconductors\cite{Nguyen_Koopmans_2018}.
    Conversely, DFT$+U$ can only achieve such improvements if the Hubbard manifold sufficiently overlaps with the frontier orbitals~\cite{himmetoglu_hubbard-corrected_2014,zhao_global_2016, kirchner-hall_extensive_2021}.
    This might explain why DFT$+U$ has been successfully applied to many Mott-Hubbard insulators, while its predictive capabilities are less reliable when applied to charge-transfer (CT) insulators.
    While in the former both the valence band maximum (VBM) and the conduction band minimum (CBM) primarily consist of localized TM $d$ states, the frontier states of the latter possess significant ligand orbital character.
    In this case, the sole correction of TM $d$ shells does not necessarily target all major sources of SIE. Additionally, the representation of the hybridized orbitals through atomic-like Hubbard projectors can be deceptive.
    
    Some authors have explored including ligand orbitals in the on-site Hubbard manifold, for example applying $U$ corrections to oxygen and S-$p$ shells of TM oxides, perovskites and sulfides \cite{moore_high-throughput_2022, may_improved_2020, orhan_first-principles_2020, kirchner-hall_extensive_2021, Gelin:2023}.
    The extended DFT+$U$+$V$ framework\cite{leiria_campo_jr_extended_2010} was adopted for scenarios where hybridization plays an eminent role. This method augments DFT$+U$ by an inter-site Hubbard $V$ term, and so acts on combinations of projectors located on different sites, enhancing accuracy and transferability~\cite{Timrov:2020c, Ricca:2020, Floris:2020, timrov_accurate_2022, kulik_transition-metal_2011, mahajan_importance_2021, mahajan_pivotal_2022, Timrov:2023, Binci:2023, Bonfa:2023, Lee:2020, Gebreyesus:2023}.
    Finally, many open-shell systems require unlike-spin terms such as Hund's $J$. In DFT+$U$+$J$, these terms introduce anisotropy in the Hubbard correction with respect to the spin channels, while also reducing the effective value of the $U$ parameter~\cite{himmetoglu_hubbard-corrected_2014,shishkin_challenges_2017,shishkin_evaluation_2021,moore_high-throughput_2022, orhan_first-principles_2020}.
    On the other hand, it is unlikely that multi-reference configurations, as common in open-shell systems, might be described well with any $+U$ or $U$+$J$ correction.
    
    What is striking about the aforementioned approaches is that the $d$ shell Hubbard manifold is either \textit{extended} by additional states ($U$ on ligands or $V$ on molecular orbitals) or internally \textit{rebalanced} (DFT+$U$+$J$). But what if the practice of including the entire $d$ shell is, in itself, problematic?
    Recent work by \citeauthor{mariano_biased_2020} suggests that this might indeed occur in some cases. Their study revealed that applying $U$ corrections to the $d$ shells of \ce{Fe} in strong-field Fe(II) hexacomplexes leads to a spurious suppression of LS states due to what appears to be an over-correction of hybridized $e_g$ orbitals \cite{mariano_biased_2020,mariano_improved_2021}.
    Furthermore, the use of \textit{ab initio} Hubbard $U$ parameters derived from LR-cDFT led to a significant decrease in the overall accuracy of the spin-state energies when compared to empirical $U$ values.
    These and other observations suggest a critical analysis of the current practice of how (shell-averaged) Hubbard $U$ corrections are applied to DFT. After all, the correction of all magnetic quantum orbitals within a given shell using the same scalar $U$ parameter is inherently a simplistic approximation~\cite{oregan_subspace_2012}.
    
    Therefore, it is worthwhile to investigate whether departing from the shell-averaged approximation can improve DFT$+U$ calculations concerning energetic, structural, and magnetic properties.
    This possibility has been explored in a few works. For example, \citeauthor{solovyev_t_1996} showed that a selective application of $U$ corrections to the $t_{2g}$ manifolds of \ce{La\textit{M}O3} perovskites (\ce{\textit{M}=Ti-Cu}) significantly improves the agreement of band gaps and magnetic orderings with experiments compared to both uncorrected LDA and shell-averaged LDA$+U$~\cite{solovyev_t_1996}. The authors argue that $e_g$ electrons are reasonably described by the uncorrected LDA functional because of their ``itinerant`` behavior that arises due to the strong $\sigma$ overlap between \ce{O}-$2p$ and \ce{M}-$e_g$ orbitals.
    \citeauthor{pickett_reformulation_1998} adopted the cDFT approach to compute the Hubbard parameters of \ce{Fe} in bulk \ce{FeO} as a matrix ($\mathbf{U}$) with $U$ values specific to the $t_{2g}$ and $e_g$ manifolds, respectively~\cite{pickett_reformulation_1998}.
    Using a different scheme, orbital-resolved Hubbard parameters were obtained by mapping shell-averaged $U$ and $J$ parameters onto orbital-dependent interaction and exchange matrices $U_{mm'}$ and $J_{mm'}$ using atomic Slater integrals and Gaunt's numbers \cite{czyzyk_local-density_1994,ylvisaker_anisotropy_2009,kasinathan_pressure-driven_2007,tompsett_importance_2012}.
    Other works focused on the spin-dependence of $U$, finding that the spin-resolved on-site parameters can be pivotal for a physical description of magnetic systems \cite{linscott_role_2018,shishkin_dft_2019,haddadiOnsiteIntersiteHubbard2023}.
    
    Despite these early efforts, orbital-resolved Hubbard $U$ parameters have not gained widespread use in the DFT community. 
    This is surprising in light of the fact that orbital-specific Hubbard manifolds are quite common in the dynamical mean-field theory (DMFT) community \cite{georges_dynamical_1996,vaugier_hubbard_2012,hampel_correlation-induced_2021, merkelCalculationScreenedCoulomb2023}.
    In the latter, Hubbard $U$ parameters are routinely computed for manifolds that range from groups of orbitals (e.g., $t_{2g}$ models) to combinations of multiple shells localized on different atoms (e.g., $d-p$ models)~\cite{vaugier_hubbard_2012} using cRPA in conjunction with various sets of Wannier projector functions that encompass the specific manifold of interest~\cite{Springer:1998,Kotani:2000}.
    To date, only a few publicly available DFT codes incorporate this capability. For example, DFT codes that use Wannier function-based Hubbard projectors (e.g., \textsc{ONETEP}\cite{prentice_span_2020}) indirectly facilitate orbital-resolved DFT$+U$ calculations as the Wannier functions can be chosen to only represent the desired subset of orbitals, e.g., $t_{2g}$\cite{oregan_projector_2010}.
   
    In this paper we present a user-friendly yet general implementation of orbital-resolved DFT$+U$ that works with any kind of Hubbard projector.
    The numerical values of the necessary parameters are extracted from first principles using an orbital-resolved LR-cDFT approach \cite{cococcioni_linear_2005}. 
    We benchmark the orbital-resolved scheme by carrying out calculations of bulk pyrite (\ce{FeS2}), bulk pyrolusite (\ce{\beta-MnO2}) and six \ce{Fe(II)} molecular hexacomplexes of varying ligand strength.
    For all of these strongly covalent compounds, the refined approach leads to a substantial improvement in the prediction of structural and energetic properties, aligning more closely with experimental observations than conventional DFT$+U$ or even DFT${+U \! + \! V}$. Moreover, we observe that the orbital-resolved $U$ parameters are considerably smaller, by up to 80\%, than their corresponding shell-averaged counterparts.  
 
    The remainder of the paper is organized as follows. In Section \ref{sec:theory}, we describe the transition from the customary, shell-averaged implementations of the DFT$+U$ energy functional and the LR-cDFT approach to their generalized, orbital-resolved forms.
    Subsequently, Section~\ref{sec:comput_details} briefly lists the relevant technical details of the calculations, whose results are presented and discussed in Section \ref{sec:results_and_discussion}.
    Section~\ref{sec:final_remarks} is dedicated to the relationship between the orbital-resolved (on-site) Hubbard $U$ presented in this work and the (inter-site) $V$ terms of Ref.~\citenum{leiria_campo_jr_extended_2010}. Finally, we summarize the main conclusions in Section~\ref{sec:conclusions}. 

\section{DFT\texorpdfstring{$+U$}{+U}: orbital-resolved Hubbard parameters}
\label{sec:theory}
    \subsection{The energy functional}
    As this work focuses on the on-site $U$ term of Hubbard-corrected DFT, we start from the widespread shell-averaged and rotationally-invariant formulation of DFT+$U$ by \citeauthor{dudarev_electron-energy-loss_1998}\cite{dudarev_electron-energy-loss_1998}. Written in a way that allows for the simultaneous correction of multiple subshells on the same atom, the energy functional reads\cite{kulik_perspective_2015}:
    \begin{equation}
        \label{eq:hubbardU-1-Eu}
        E_{\mathrm{DFT}+U} = E_{\mathrm{DFT}} + E_{U} =  E_{\mathrm{DFT}} + \sum_{I,\sigma}\sum_{nl} \frac{U^I_{nl}}{2} \, \mathrm{Tr} \left[\mathbf{n}^{I\sigma}_{nl}(1-\mathbf{n}^{I\sigma}_{nl}) \right] \,,
    \end{equation}
    where $E_\mathrm{DFT}$ is the DFT total energy computed with standard (semi-)local xc functionals, $U^I_{nl}$ is an effective on-site Hubbard parameter, and $\mathbf{n}^{I\sigma}_{nl}$ is the orbital occupation matrix. 
    The summation over the principal ($n$) and orbital ($l$) quantum numbers implies that Hubbard $U$ corrections can be applied simultaneously to multiple subshells of the same atom\cite{leiria_campo_jr_extended_2010,kirchner-hall_extensive_2021}, although it usually suffices to treat the valence shell alone. For the sake of clarity, we henceforth omit the index $nl$, assuming that the Hubbard manifolds consist of, at most, one subshell per atom.
    The orbital occupation matrix defines how the Hubbard correction is applied. For a given spin $\sigma$, its elements are computed by projecting the valence KS wave functions onto the Hubbard manifold of an atom $I$:
    \begin{equation}
       \label{eq:occup_general}
       {n}^{I\sigma}_{mm'} = \sum_{\mathbf{k},\nu} f_{\mathbf{k},\nu}^{\sigma} \langle \psi_{\mathbf{k},\nu}^{\sigma}|\hat{P}^I_{m'm}|\psi_{\mathbf{k},\nu}^{\sigma} \rangle \,,
    \end{equation}
    where $m$ and $m'$ are the magnetic quantum numbers of the Hubbard manifold, $\nu$ represents the band labels of the KS wave functions, $\mathbf{k}$ indicates points in the first Brillouin zone, and $f_{\mathbf{k},\nu}^{\sigma}$ are occupations of the KS wave functions $\psi^{\sigma}_{\mathbf{k},\nu}$.
    If atomic-like orbitals ($\varphi^I_{m}$) are used as projector functions (\textit{vide infra}), one may define $\hat{P}^I_{m'm} = |\varphi^I_{m'} \rangle \langle\varphi^I_{m}|$. Note, however, that if plane-wave basis sets are used, the expression of $\hat{P}$ also depends on the type of the pseudopotential~\cite{timrov_ultrasoft_2021}.
    Applying Hubbard $U$ corrections also modifies the KS potential of the target orbitals according to
    \begin{equation}
        \label{eq:hubbard-potential-u}
        \hat{V}_{U}^\sigma = \sum_{I} U^I \sum_{mm'} \left(\frac{\delta_{mm'}}{2} -  n^{I\sigma}_{mm'} \right)|\varphi^I_m \rangle \langle \varphi^I_{m'}| ,
    \end{equation}
    where $\delta_{mm'}$ is the Kronecker delta. It is evident from Eq. \eqref{eq:hubbard-potential-u} that Hubbard $U$ corrections exert a stabilizing influence on fully occupied orbitals, reducing their KS potential by up to $U/2$, while producing the opposite effect on empty orbitals.
    
    Provided a correct normalization of the Hubbard projector functions, the eigenvalues of $n^{I\sigma}_{mm'}$ express the occupation of the $2l+1$ orbitals with numbers between 0 (fully empty) and 1 (fully occupied). They are obtained by solving the eigenvalue problem \cite{cococcioni_linear_2005}
    \begin{equation}
         \label{eq:lambda_def}
        \mathbf{n}^{I\sigma} \mathbf{v}_i^{I\sigma} = \lambda_{i}^{I\sigma} {\mathbf{v}}_i^{I\sigma}, 
    \end{equation}
    where $\lambda_{i}^{I\sigma}$ and $\mathbf{v}_i^{I\sigma}$ are the eigenvalues and eigenvectors, respectively, and $i$ is a dummy index running from 1 to $2l+1$. 
    This allows us to re-write the $E_U$ contribution of Eq.~\eqref{eq:hubbardU-1-Eu} more concisely in terms of these eigenvalues:
    \begin{equation}
        \label{eq:hubbardU}
        E_U = \sum_{I,\sigma} \frac{U^I}{2} \sum_{i}^{2l+1} \lambda_{i}^{I\sigma}(1-\lambda_{i}^{I\sigma}) \,.
    \end{equation}
    Within this diagonal representation, Eq.~\eqref{eq:hubbardU} can be readily generalized to become an orbital-resolved correction:
    \begin{equation}
        \label{eq:hubbardUm}
        E_U = \sum_{I,\sigma} \sum_{i}^{2l+1} \frac{U_i^I}{2} \lambda_{i}^{I\sigma}(1-\lambda_{i}^{I\sigma}) \,,
    \end{equation}
    Here, $U_i^I$ is now an effective on-site Hubbard parameter specific to the $i$th orbital of the Hubbard manifold localized on atom $I$.
   
    Similar generalizations of the shell-averaged DFT+$U$ functional were already postulated by \citeauthor{pickett_reformulation_1998} and used by \citeauthor{solovyev_t_1996}, although these early works adopted DFT$+U$ energy functionals different from \citeauthor{dudarev_electron-energy-loss_1998}'s formulation.
    
    In practical calculations, not every orbital requires a distinct $U_i^I$, and it might suffice to distinguish between the irreducible representations that follow from local point group symmetry.
    For instance, considering the $d$ shell of an octahedrally coordinated atom with local $O_h$ point group symmetry, Eq.~\eqref{eq:hubbardUm} can be written as
    \begin{equation}
        \label{eq:hubbardUm2} 
        E_U = \sum_{I,\sigma} \frac{U_{t_{2g}}^I}{2} \sum_{i \in \{t_{2g}\}} \lambda_{i}^{I\sigma}(1-\lambda_{i}^{I\sigma})
        + \sum_{I,\sigma} \frac{U_{e_{g}}^I}{2} \sum_{i \in \{e_{g}\}} \lambda_{i}^{I\sigma}(1-\lambda_{i}^{I\sigma}) \,,
    \end{equation}
    where $U_{t_{2g}}^I$ and $U_{e_{g}}^I$ are the Hubbard parameters for the $t_{2g}$ and $e_{g}$ orbitals, respectively, and $i \in \{t_{2g}\}$ means that $i$ runs over the orbital indices of the $t_{2g}$ subshell, while $i \in \{e_{g}\}$ means that $i$ runs over the orbital indices of the $e_{g}$ subshell.
    It is also possible to selectively exclude specific orbitals from receiving Hubbard corrections by setting $U^I_i = 0$. 

    The calculation of the Hubbard potential (Eq. \eqref{eq:hubbard-potential-u}) is also performed in the diagonal representation by rotating the atomic-like orbitals $\varphi^I_m$ from the global to the local coordinate system using the eigenvectors $\mathbf{v}_i^{I\sigma}$. However, after computing the orbital-resolved contributions to the Hubbard potential $\hat{V_{U}}$, we perform a backrotation into the non-diagonal representation as this allows to use existing implementations of density mixing as well as the calculation of forces and stresses with no further adaption.
    
    A crucial aspect for a successful application of Hubbard corrections lies in finding a suitable projector function for the target manifold, as this controls the occupation eigenvalues $\lambda_{i}^{I\sigma}$ (Eq.~\eqref{eq:occup_general}) which govern the corrective Hubbard energy. The next subsection offers a concise introduction to the prevailing and commonly utilized approaches.

    \subsection{Hubbard projector functions}
        \label{sec:Hub_projector_functions}
    There are many ways to define Hubbard projector functions within the DFT+$U$ approach, and the reader is referred to Ref.~\citenum{timrov_pulay_2020} and references therein for a more comprehensive overview.
    For electronic-structure codes employing a localized basis set, natural choices for Hubbard occupations are either Mulliken or L\"owdin population matrices~\cite{kulik_adapting_2016}.
    In contrast, codes based on plane-wave basis sets often use atomic orbitals as projectors~\cite{cococcioni_linear_2005, rohrbach_molecular_2004}, whose localized functions are parameterized with free-atom calculations and then stored in the pseudopotentials.
    During the generation of pseudopotentials, the atomic orbitals are chosen to be orthonormal to all other orbitals centered on the same atom. However, this cannot guarantee orthogonality to orbitals localized on other sites during practical calculations. 
    Hence, these projector functions are referred to as nonorthogonalized atomic orbitals (NAO)~\cite{mahajan_importance_2021}.
    While straightforward to implement and use, NAO may display spatially extended `tails', potentially resulting in the same domain being tackled twice by Hubbard corrections if the atomic orbitals of two neighboring atoms overlap.
    
    Truncation spheres~\cite{rohrbach_molecular_2004} provide a means of ruling out such double counting by cutting off possible tails.~\footnote{Note that this ``double counting" must not be confused with the double counting term in DFT+$U$ energy functionals derived from the Hubbard model.}
    However, the radius of these spheres represents an additional parameter that affects the obtained occupation numbers~\cite{nawa_scaled_2018}. The value of this parameter varies depending on the code utilized and may require manual adjustment (e.g., muffin-tin radius in the linearized augmented plane-wave approach) or can be embedded within the pseudopotential (typical of projector-augmented-wave (PAW) pseudopotentials).
    
    An alternative, parameter-free approach to circumvent truncation involves orthogonalizing NAO across all atomic sites, e.g., using L\"owdin's scheme~\cite{lowdin_nonorthogonality_1950}, thus transforming them into orthogonalized atomic orbitals (OAO)~\cite{mahajan_importance_2021,timrov_pulay_2020}. 
    The resulting inter-site orthogonality clears the overlap of projectors on different sites~\cite{timrov_pulay_2020} and even accounts for possible hybridization between them, albeit to a limited extent. Several benchmark studies have shown that OAO projectors consistently outperform NAO projectors with respect to structural, electronic and spectral properties\cite{mahajan_importance_2021,mahajan_pivotal_2022,kirchner-hall_extensive_2021}.
    
    Finally, Wannier functions are also viable Hubbard projector functions~\cite{Nakamura:2006, Korotin:2012, Novoselov:2015}. Specifically, maximally localized Wannier functions (MLWF)~\cite{Marzari:1997, Marzari:2012} can separate manifolds in a system-specific fashion~\cite{Qiao:2023}, and can serve as effective Hubbard projector functions within the generalized DFT+$U$ framework presented in Eq.~\eqref{eq:hubbardUm}. Despite the extensive use of Wannier functions, their current adoption in DFT+$U$ calculations remains limited~\cite{tesch_hubbard_2022,sakuma_first-principles_2013,oregan_projector_2010}, possibly due to the non-trivial steps of finding an appropriate starting guess and disentangling overlapping bands during wannierization. Moreover, features relevant for practical studies such as the calculation of forces and stresses are cumbersome to implement.
    
    \subsection{LR-cDFT to compute orbital-resolved \texorpdfstring{$U$}{U} parameters}
    \label{subsec:lr-cdft}   
    LR-cDFT is based on using the DFT$+U$ energy functional to (heuristically) restore piecewise-linearity for the Hubbard manifold in (semi-)local DFT functionals suffering from electron self-interaction\cite{cococcioni_linear_2005,kulikDensityFunctionalTheory2006}.
    An important manifestation of the latter are so-called fractional charge errors (FCE), which are spurious (usually convex) deviations from linearity of $E_{\mathrm{DFT}}$ with respect to fractional addition or removal of charge~\cite{zhao_global_2016}, i.e.,
    \begin{equation}
    \label{eq:global-curvature}
        \mathrm{FCE} = \frac{\partial^2 E_{\mathrm{DFT}}}{\partial q^2}, 
    \end{equation}
    where $q$ represents the charge of the system under consideration.
    Inspection of Eq.~\eqref{eq:hubbardU-1-Eu} shows that the Hubbard $U$ correction amounts to removing a quadratic term and adding a linear one, scaled by the numerical value of $U$. Note, however, that the curvature removed by Eq.~\eqref{eq:hubbardU-1-Eu} is not with respect to the total charge of the system $q$ but instead with respect to the projected \textit{local} occupation of the Hubbard manifold $\mathbf{n}^{I\sigma}$. Thus, a fundamental assumption is that the electrons in this localized Hubbard manifold are the most affected by self-interaction and can be dealt with separately.
    This interpretation makes it possible to define the value of Hubbard $U$ as the one for which the second derivative of the total energy functional becomes zero with respect to changes in the occupation of the shell~\cite{cococcioni_linear_2005},
    \begin{equation}
        \label{eq:local-curvature}
        U^I = \frac{\partial^2 E}{\partial (\Lambda^I)^2} \rvert_q  \,,
    \end{equation}
    where we now use $\Lambda^I = \sum_\sigma \sum_i^{2l+1}\lambda^{I\sigma}_i$ to define the local occupation of the Hubbard manifold.
    Because a direct control of orbital occupations is not tractable in codes that obtain them as output quantities, Lagrange multipliers $\alpha$ are introduced to linearly shift the potential of the Hubbard manifold and thus indirectly control $\Lambda^I$ (see Refs.~\citenum{pickett_reformulation_1998,cococcioniLDAStudySelected2002,cococcioni_linear_2005} for the derivation).
    Then, two response matrices are defined using finite differences:
    \begin{equation}
        \label{eq:lr-chi}
        (\chi)^{IJ} = \frac{\Delta \Lambda^I}{\Delta \alpha^J} \,, \quad  
        (\chi_0)^{IJ} = \frac{\Delta \Lambda^I_0}{\Delta \alpha^J} \, ,
    \end{equation}
    where $\chi$ represents the self-consistent, screened response of the manifold, whereas $\chi_0$ is due to the non-interacting, unscreened response coming from the re-hybridization of the atomic orbital projectors that results from the perturbation. In many plane-wave codes, these latter non-interacting orbital occupations $\Lambda^I_0$ can be obtained from the first iteration in the self-consistent cycle of the perturbative calculation.
    Since the response $\chi_0$ is unrelated to electron self-interaction, it must be subtracted from the interacting (i.e., screened) response when computing $U$:
    \begin{equation}
        U^I = \left( \chi_{0}^{-1} - \chi^{-1} \right)^{II} \,.
        \label{eq:u-lr-chi}
    \end{equation}
    In practice, the response functions are obtained by either applying multiple small (positive and negative) perturbations to the shells of interest of a converged ground state~\cite{cococcioni_linear_2005} (with periodic systems requiring a supercell approach to avoid interactions between a perturbed Hubbard manifold and its periodic images) or from DFPT~\cite{timrov_hubbard_2018,timrov_hp_2022} using the response to monochromatic perturbations in a primitive cell.
    
    Orbital-resolved Hubbard $U$ parameters can be evaluated using the formalism of Eqs.~\eqref{eq:lr-chi} and \eqref{eq:u-lr-chi} by adapting the definition of the total occupation of the Hubbard manifold. 
    In the most general case, every magnetic quantum orbital $i$ of the $nl$ subshell can acquire an individual Hubbard parameter. The occupation of such a manifold is given by
    \begin{equation}
        \label{eq:Lambda-um}
        \Lambda^I_{i} = \sum_\sigma \lambda^{I\sigma}_{i} \, .
    \end{equation}
    With this, the elements of the response matrices (Eq.~\eqref{eq:lr-chi}) can be redefined as
    \begin{equation}
        \label{eq:lr-chi-full}
        (\chi)^{IJ}_{ij} = \frac{\Delta \Lambda^I_{i}}{\Delta \alpha^J_{j}} \,, \quad  
        (\chi_0)^{IJ}_{ij} = \frac{\Delta \Lambda^I_{i,0}}{\Delta \alpha^J_{j}} \, ,
    \end{equation}
    while the expression for the orbital-resolved on-site Hubbard $U$ parameters becomes
    \begin{equation}
        U^I_{ij} = \left( \chi_{0}^{-1} - \chi^{-1} \right)^{II}_{ij} \,.
        \label{eq:um-lr-chi}
    \end{equation}
    The generalized formalism of Eq.~\eqref{eq:um-lr-chi} not only accommodates orbital-resolved on-site $U$ parameters, but allows for the determination of inter-site parameters ($U^{IJ}_{ij}$ with $I \neq J$, or, following the nomenclature of DFT${+U \! + \! V}$, $V^{IJ}_{ij}$), as well as on-site inter-orbital parameters ($U^{II}_{ij}$ for $i\neq j$).  Nevertheless, the primary emphasis of this study lies in investigating the on-site intra-manifold parameters $U^{II}_{ii} \equiv U^I_i$.

    For the sake of illustration, let us consider a \textit{d} shell of an atom exhibiting local $O_h$ symmetry. Also, we assume that this atom be the only Hubbard atom in the system, thereby enabling us to neglect inter-site responses and to drop the superscript $I$.
    Then, one can define an orbital-resolved Hubbard matrix of size $2 \times 2$ that reads
    \begin{equation}
        \label{eq:3d-response} 
        \begin{pmatrix}
                U_{t_{2g},t_{2g}} & U_{e_{g},t_{2g}} \\
                U_{t_{2g},e_{g}}  & U_{e_{g},e_{g}}
        \end{pmatrix}
        \\
        =
        \begin{pmatrix}
                (\chi_0)_{t_{2g},t_{2g}} & (\chi_0)_{e_{g},t_{2g}} \\
                (\chi_0)_{t_{2g},e_{g}}  & (\chi_0)_{e_{g},e_{g}}
        \end{pmatrix}^{-1}
         - 
        \begin{pmatrix}
            (\chi)_{t_{2g},t_{2g}} & (\chi)_{e_{g},t_{2g}} \\
            (\chi)_{t_{2g},e_{g}}  & (\chi)_{e_{g},e_{g}}
        \end{pmatrix}^{-1} ,
    \end{equation}
    where the indices represent the response in occupations of manifold $a$ to perturbation of manifold $b$, 
    $(\chi)_{a,b} = \frac{\Delta \Lambda_{a}}{\Delta \alpha_{b}}$, for instance $(\chi)_{e_{g},t_{2g}} = \frac{\Delta \Lambda_{e_g}}{\Delta \alpha_{t_{2g}}}$.
    \citeauthor{pickett_reformulation_1998} showed that the orbital-resolved matrix elements of Eq.~\eqref{eq:3d-response} are related to the shell-averaged Hubbard parameter $U$ through a sum rule~\cite{pickett_reformulation_1998}. A slightly adapted form of this rule that accounts for the role of the non-interacting response reads\footnote{\citeauthor{pickett_reformulation_1998} define the $d$ shell-averaged Hubbard parameter (in \textit{their} notation) as $U_{dd}=\left [ \sum_{a,b=t_{2g},eg} (\chi)_{a,b} \right ]^{-1}$, whereas in this work we use the definition of Eq. \eqref{eq:pickett} due to the non-interacting (bare) response.}
    \begin{eqnarray}
        U & = & \left[\sum_{a,b=t_{2g},e_g} (\chi_0)_{a,b}\right]^{-1} -  \left[\sum_{a,b=t_{2g},e_g} (\chi)_{a,b}\right]^{-1} .
        \label{eq:pickett}
    \end{eqnarray}

    The lesson to learn from Eq.~\eqref{eq:pickett} lies in the off-diagonal values of $\chi$, whose physical implication consists in intra-shell screening (in this example $e_g \leftrightarrow t_{2g}$). 
    Since the sign of these off-diagonal values is normally opposite to that of the diagonal ones \footnote{This results from a physical necessity: if a perturbation to the potential of a manifold $a$ causes the occupation of $a$ to increase, the electrons must be borrowed from another manifold, whose orbital occupation will naturally decrease, since the global charge is kept constant in LR-cDFT.}, they diminish the contribution of the term $\sum_{ab}(\chi)_{a,b}$, which may substantially increase $U$ values upon computation of the inverse. Note that $\chi_0$ is not affected by this, because the off-diagonal elements of the unscreened response are zero by definition (except for numerical noise).
    In the extreme scenario where perturbations are exclusively screened within the same shell, the sum over the off-diagonal elements equals the trace of $\chi$, causing $\det(\chi)$ to approach zero, leading to $U \to \infty$.
    Consequently, if the current definition of the response matrices is used to compute orbital-resolved $U$ parameters, the determinant required for the full inversion of $\chi$ inadvertently reintroduces the shell-averaging of the response. This is evident in the findings of the aforementioned study by \citeauthor{pickett_reformulation_1998}, where the computed $U_{t_{2g}}$ ($\equiv U_{t_{2g},t_{2g}}$) displayed only minimal differences compared to $U_{e_g}$ ($\equiv U_{e_g,e_g}$) and the shell-averaged $U$ \cite{pickett_reformulation_1998}. 

    In order to derive on-site $U$ parameters that explicitly incorporate intra-shell screening effects, it is necessary to set the off-diagonal matrix elements of $\chi$ and $\chi_0$ to zero before computing their inverses.
    This scheme was employed by \citeauthor{linscott_role_2018}, who refer to it as pointwise (``$1\times1$'') inversion, in order to compute screened and spin-resolved Hubbard parameters for metal aquo complexes\cite{linscott_role_2018}.
    \citeauthor{solovyev_t_1996} also applied this strategy, albeit implicitly, to obtain screened $U$ parameters specific to $t_{2g}$ in perovskites \cite{solovyev_t_1996}. 
    Although the removal of off-diagonal elements seems a drastic approximation, it allows for a more tailored definition of Hubbard manifolds compared to the conventional DFT$+U$ approach. The latter assumes (without proof) that ligand orbitals account for the majority of screening while the role of intra-shell interactions is neglected. 
    In fact, several cRPA studies suggest that the opposite is true, showing that intra-shell screening can be as significant, or even more so, as inter-shell screening \cite{vaugier_hubbard_2012,hampel_correlation-induced_2021}.
    An important caveat to the LR-cDFT approach that also affects the orbital-resolved form presented here is that unphysical $U$ values may result when it is applied to fully occupied manifolds. This is a well-known limitation~\cite{kulik_systematic_2010,Yu:2014} that follows from the fact that the response of a deep-lying state to a relatively small perturbation ($\approx0.05\,$eV) is often on the same order of magnitude as the numerical noise, leading to instabilities during the inversion of the response matrices.

\section{Computational details}
\label{sec:comput_details}
    All calculations are carried out using the \QE distribution~\cite{Giannozzi:2009, Giannozzi:2017, Giannozzi:2020}. We have incorporated the capability to utilize and determine orbital-resolved Hubbard $U$ parameters with the \texttt{pw.x} code and will make this accessible in a future release. Structure and isosurface plots for pyrite (\ce{FeS2}) and pyrolusite (\ce{\beta-MnO2}) are generated using \texttt{VESTA}\cite{VESTA_2011}.
    Unless stated otherwise, all systems are structurally optimized using the Broyden-Fletcher-Goldfarb-Shanno (BFGS) algorithm~\cite{Fletcher:1987} with convergence thresholds of $10^{-4}\,$Ry, $10^{-3}\,$Ry/Bohr, and $0.5\,$kbar for the total energy, forces, and pressure, respectively. KS wavefunctions (charge density) are expanded in plane waves up to a kinetic-energy cutoff of 90\,Ry (1080\,Ry) using PBE~\cite{perdew_generalized_1996} pseudopotentials for pyrite and the \ce{Fe(II)} molecular complexes, and PBEsol pseudopotentials for \ce{\beta-MnO2} \footnote{
    We used \texttt{Fe.pbe-spn-kjpaw\_psl.0.2.1.UPF}~\cite{kucukbenli_projector_2014}, \texttt{s\_pbe\_v1.4.uspp.F.UPF}\cite{garrity_pseudopotentials_2014}, \texttt{O.pbe-n-kjpaw\_psl.0.1.UPF}\cite{kucukbenli_projector_2014}, \texttt{H.pbe-rrkjus\_psl.1.0.0.UPF}\cite{dal_corso_pseudopotentials_2014}, \texttt{P.pbe-n-rrkjus\_psl.1.0.0.UPF}\cite{dal_corso_pseudopotentials_2014}, \texttt{C.pbe-n-kjpaw\_psl.1.0.0.UPF}\cite{dal_corso_pseudopotentials_2014}, \texttt{N.pbe-n-radius\_5.UPF} (THEOS pseudo) and \texttt{mn\_pbesol\_v1.5.uspp.F.UPF}\cite{garrity_pseudopotentials_2014} and \texttt{O.pbesol-n-kjpaw\_psl.0.1.UPF}\cite{kucukbenli_projector_2014} for \ce{\beta-MnO2}.}
    taken from the SSSP Precision library v.\,1.1.2~\cite{lejaeghere_reproducibility_2016,prandini_precision_2018}. 
    The projected density of states (PDOS) is obtained using a Gaussian smearing with a broadening parameter of $0.02\,$Ry and employing the \textit{diag\_basis} feature of the \texttt{projwfc.x} code, which projects the wavefunctions onto the eigenstates of the occupation matrix rather than using unrotated atomic orbital projectors \cite{mahajan_importance_2021}. This allows for a clear distinction between $t_g$/$t_{2g}$ and $e_g$ states, regardless of the orientation of the global coordinate system.
    \ce{Fe(II)} molecular complexes are simulated at a fixed $+2$ charge state in cubic boxes with an edge length of $15\,$\AA~and using only the $\Gamma$ point to sample the Brillouin zone. The total magnetization of the molecular complexes is always fixed to either $4.0\,\mu_{\mathrm{B}}$ or $0.0\,\mu_{\mathrm{B}}$ in order to compute high-spin (HS) and low-spin (LS) configurations, respectively. The starting geometries of the \ce{Fe(II)} molecular complexes are taken from the SI of Ref.~\citenum{mariano_biased_2020}, whereas experimental structures are chosen as a starting points for \ce{FeS2} and \ce{\beta-MnO2}.
    The Brillouin zones of \ce{FeS2} and \ce{\beta-MnO2} are sampled with uniform $\Gamma$-centered Monkhorst-Pack meshes of sizes $9\times9\times9$ and $4\times4\times6$, respectively.

    We use L\"owdin-orthogonalized atomic orbitals as Hubbard projectors (OAO)~\cite{lowdin_nonorthogonality_1950,timrov_pulay_2020} for all DFT+$U$ calculations including those with orbital-resolved $U$ parameters.
    Shell-averaged Hubbard parameters are evaluated using the DFPT implementation~\cite{timrov_hubbard_2018, timrov_ultrasoft_2021} of the \textsc{HP} code (\texttt{hp.x})~\cite{timrov_accurate_2022} included in \QE, employing $\mathbf{q}$ point meshes of size $2\times 2\times2$ for \ce{FeS2} and \ce{\beta-MnO2}, and $1\times 1\times1$ for the \ce{Fe(II)} molecular complexes, respectively.
    We compute orbital-resolved $U$ parameters according to the LR-cDFT approach described in Section \ref{sec:theory} by applying perturbations of $\alpha=[-0.05,0.05]\,$eV to the manifold of interest and recording the non-interacting and interacting responses of the orbital occupations. To avoid interactions of perturbations with their periodic images, the calculations of \ce{FeS2} and \ce{\beta-MnO2} are conducted in a $2\times2\times2$ supercell containing 96 atoms. We emphasize that DFPT and LR-cDFT are equivalent by construction, and therefore yield the same Hubbard parameters when applied to identical systems~\cite{timrov_hubbard_2018}.
    As the orbital response depends on the system's electronic structure, calculated Hubbard parameters may vary significantly when transitioning from a PBE/PBEsol ground state to a PBE/PBEsol$+U$ one. 
    Therefore, to achieve self-consistency of the computed $U$ values, we employ an iterative procedure that consists of structural optimizations and subsequent perturbative calculations~\cite{timrov_ultrasoft_2021,shishkin_self-consistent_2016,moynihan_self-consistent_2017}. This procedure is repeated until the difference between the input and output parameters falls below the predefined threshold of $\sim 0.1$\,eV. The resulting Hubbard parameters are reported in Section~\ref{sec:results_and_discussion}.


\section{Results and discussion}
    \label{sec:results_and_discussion}
    \subsection{Pyrite (\texorpdfstring{\ce{FeS2}}{FeS2})}
        \subsubsection{Challenging theoretical description}
            \begin{figure}[ht]
                \includegraphics[width=3.33in]{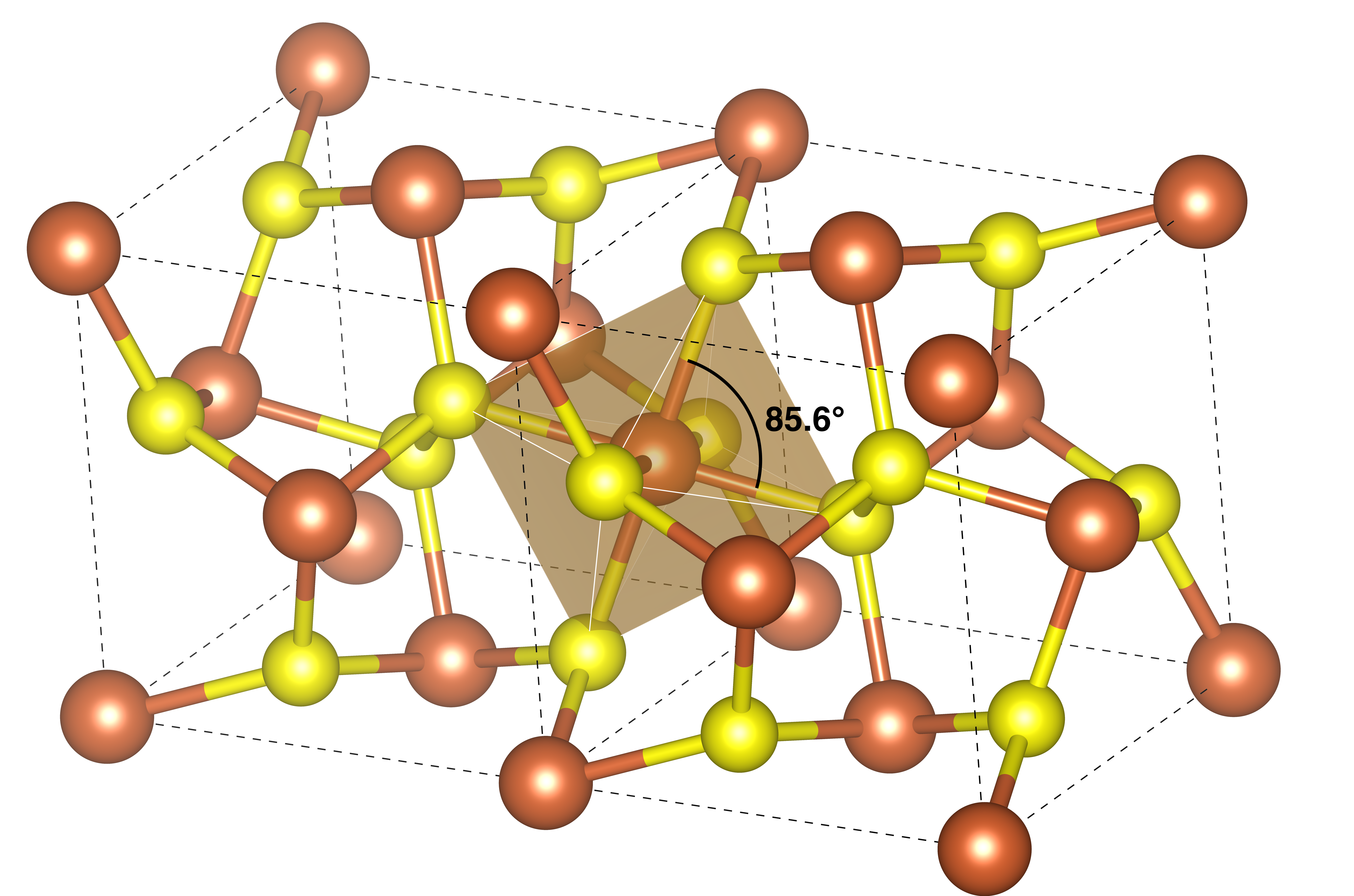}
                \caption{The experimental crystal structure of pyrite (\ce{FeS2}). \ce{Fe} (brown) is octrahedrally coordinated by \ce{S} (yellow), which also forms characteristic \ce{S-S} dimers. The octahedra are slightly distorted and display angles different from $90^{\circ}$.}
                \label{fig:pyrite_structure}
            \end{figure}
            Under normal conditions, pyrite is the stable polymorph of \ce{FeS2} and crystallizes in the cubic space group \ce{Pa$\Bar{3}$} with an experimental lattice parameter $a=5.418\,$\AA\cite{Zuniga_pyrite_2019}. The crystal structure, shown in Figure~\ref{fig:pyrite_structure}, consists of \ce{S2} dimers octahedrally coordinating \ce{Fe^{2+}} ions, which form a \textit{fcc} sublattice \cite{banjara_first-principles_2018}.
            At 0\,K, the compound is diamagnetic ($S=0$) due to the preferred LS configuration of the \ce{Fe^{2+}} ions. 
            Pyrite's natural abundance, optical band gap of $\sim$0.95\,eV, and large optical absorption coefficient make it an appealing material for photovoltaic applications. Nevertheless, despite theoretical predictions suggesting an open circuit voltage of $\sim$\,0.71\,V based on the Shockley-Queisser equations, experimental results have consistently fallen short, typically measuring values around 0.2\,V \cite{sun_first-principles_2011}.
            This discrepancy is among several reasons motivating an accurate quantum-mechanical description of pyrite's ground state.

\begin{figure}[h]
                \includegraphics[width=3.33in]{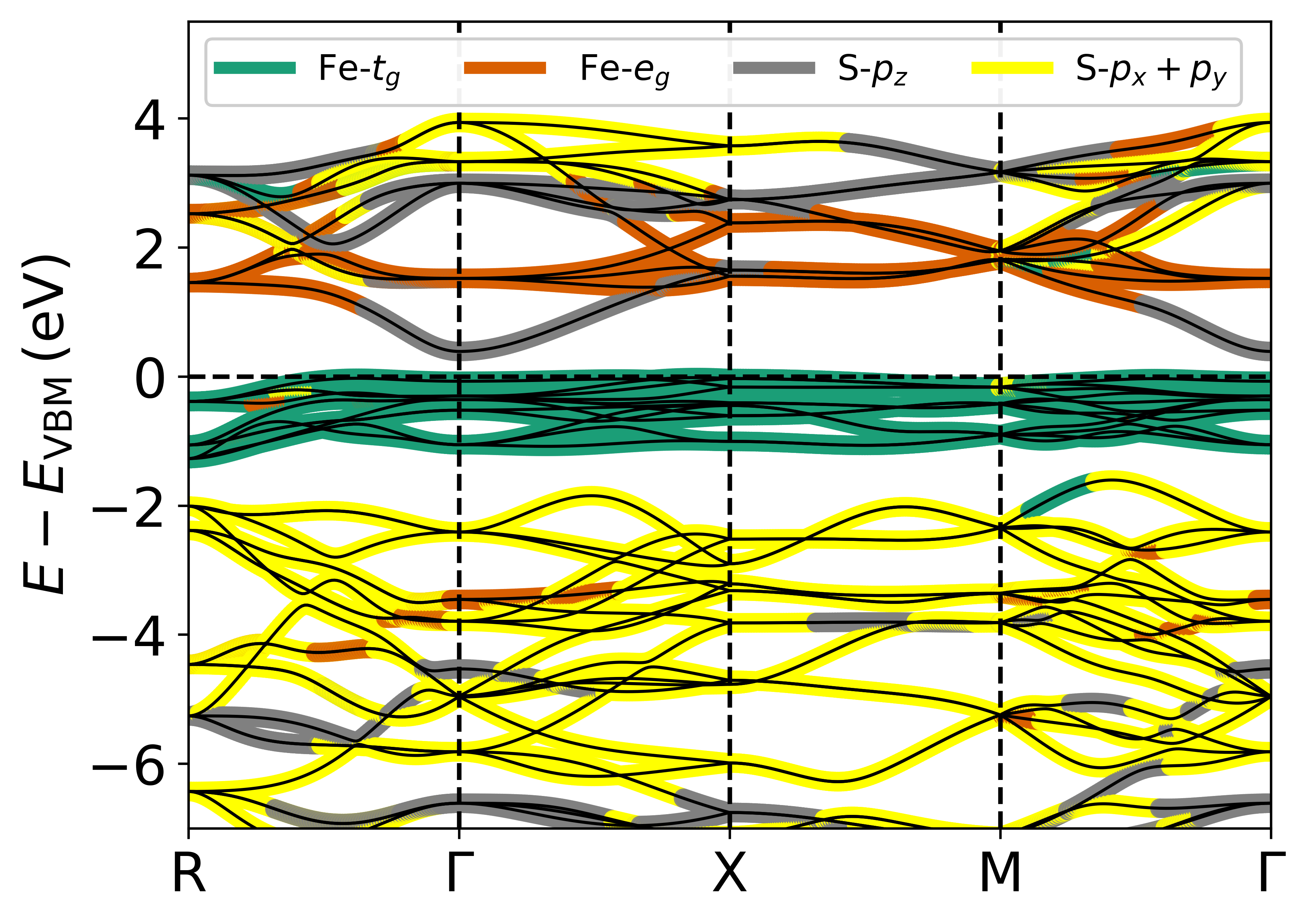}
                \caption{(Fat-)Band structure of pyrite computed using the PBE functional. The coloring indicates the most dominant contribution of the manifolds to the individual bands. The uppermost valence bands are predominantly Fe-$t_{g}$, whereas the conduction bands display a strong S-$3p_z$ character at $\Gamma$ but Fe-$e_{g}$ character elsewhere. Thus, the band gap of pyrite is not a $d$-$d$ transition, but a $d$-$p$ one.}
                \label{fig:pyrite_bandstructure}
            \end{figure}

            In early DFT studies, the LDA functional demonstrated exceptional accuracy in predicting the equilibrium volume, band gap, and relative energy levels of the \ce{S}-$3p$ bands compared to the valence band maximum (VBM) \cite{eyert_electronic_1998, banjara_first-principles_2018}. PBE\cite{perdew_generalized_1996} and AM05\cite{armiento_functional_2005} provide a qualitatively similar picture but underestimate the band gap by about 0.5\,eV and 0.75\,eV, respectively~\cite{sun_first-principles_2011}.
            \begin{figure}[ht]
                \includegraphics[width=3.33in]{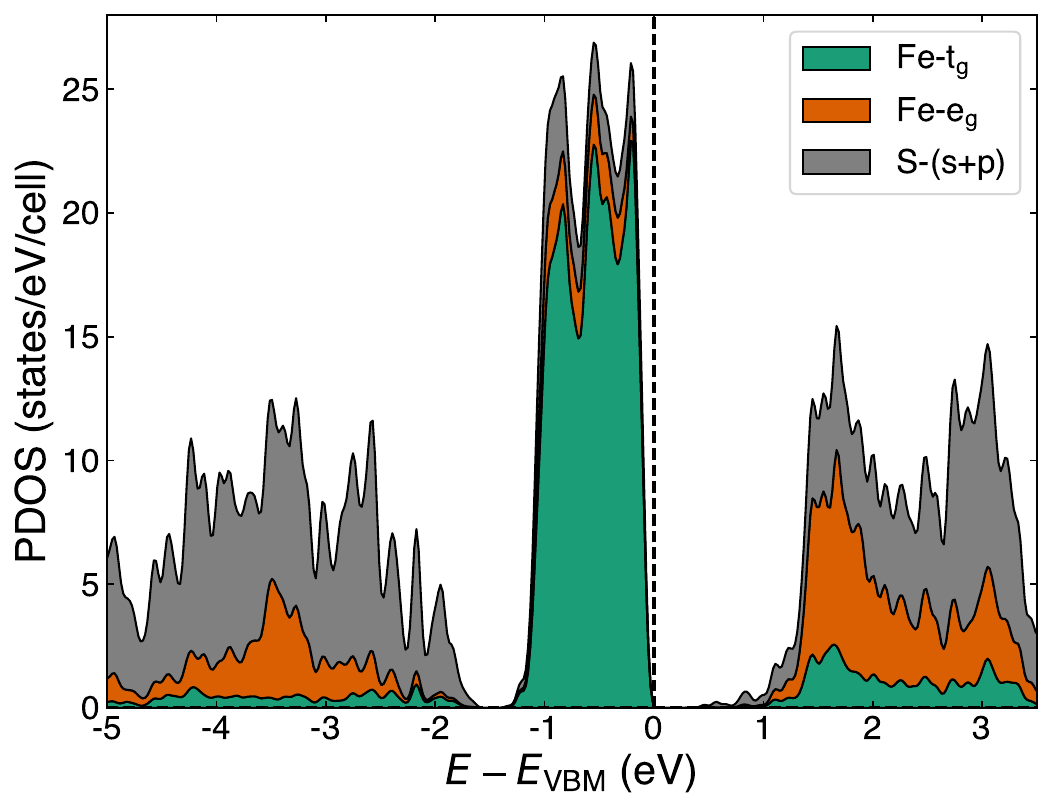}
                \caption[PDOS of pyrite.]{Stacked projected density of states (PDOS) of the nonmagnetic PBE ground-state of pyrite \ce{FeS2}. The small $t_{g}$ contributions\protect\footnotemark ~in the conduction band and the $e_g$ contributions close to the VBM are likely projection artifacts resulting from the slightly canted geometry of the \ce{FeS6} octahedra.}
                \label{fig:pyrite_pdos_pbe}
            \end{figure}
            \footnotetext{The point group symmetry of \ce{Fe} in pyrite is $T_h$ rather than $O_h$; thus, the triply degenerate irreducible representation is correctly named $t_{g}$ instead of $t_{2g}$\cite{burns_introduction_1977}.}
            At first glance, the dominant $e_g$ contributions in the conduction band (Figure~\ref{fig:pyrite_pdos_pbe}) indicate Mott-Hubbard insulation. 
            However, a more detailed analysis of the electronic band structure (Figure \ref{fig:pyrite_bandstructure}) reveals that the conduction band minimum (CBM), located at the $\Gamma$ point, is composed of a \ce{$d$-$p$ \sigma^{*}} hybrid orbital with dominant S-$3p_{z}$ contributions.
            This observation holds the key to understanding why numerous electronic structure methods, including more advanced approaches, face challenges in improving the prediction of the band gap beyond the capabilities of PBE.
            For instance, when employing $G_0W_0$, the small PBE band gap diminishes further \cite{schena_first-principles_2013}. In contrast, hybrid functionals such as PBE0\cite{perdew_rationale_1996}, HSE06\cite{heyd_hybrid_2003}, DSH\cite{cui_doubly_2018}, and M06\cite{zhao_m06_2008} overestimate its value by more than $1\,$eV \cite{zhang_accurate_2021}.
            Previous PBE$+U$ studies focused on correcting the Fe-$d$ shell using shell-averaged Hubbard parameters. For example, \citeauthor{sun_first-principles_2011} applied $U=2.0\,$eV (with PAW projectors)~\cite{sun_first-principles_2011}, whereas \citeauthor{schena_first-principles_2013} used a combination of $U=3.0\,$eV with $J=1.0\,$eV (Muffin-Tin sphere projectors)\cite{schena_first-principles_2013}.
            Notably, no set of \textit{ab initio} computed Hubbard parameters has been published for this material to date, to the best of our knowledge.
            The rationale behind the use of empirical parameters likely lies in a pronounced influence of shell-averaged $U$ corrections on the equilibrium properties, as we demonstrate hereinafter.

        \subsubsection{Double impact of shell-averaged \textit{U} corrections}
        \label{subsec:u_to_pyrite}
            We perform PBE+$U$ calculations incorporating empirical on-site Hubbard $U$ corrections ranging from $1.0$ to $5.0\,$eV and analyze their impact on the estimated band gap and on the crystal structure. 
            To differentiate between a band gap broadening caused by structural changes and one resulting directly from the influence of $U$ on the electronic structure, we perform the calculations under three different constrained conditions.
            Initially, we retain the PBE ionic structure, keeping the cell vectors and ionic positions fixed (setup~1). Subsequently, we perform another set of calculations with ionic position relaxations while maintaining a constant cell volume (setup~2). Lastly, we conduct full optimization, allowing for simultaneous adjustments of both ionic positions and cell vectors (setup~3).
            While band gaps are fundamentally outside the realm of DFT, DFT$+U$ should improve upon the performance of the uncorrected functional as long as the compound's frontier states are well-represented by the Hubbard manifold\cite{kraisler_pwl_2013, kirchner-hall_extensive_2021}. For this case, one can argue that the Hubbard $U$ corrections (locally) act in the spirit of a Koopmans-compliant functional\cite{borghiKoopmanscompliantFunctionalsTheir2014}.
            
\begin{figure}[t]
                \includegraphics[width=2.5in]{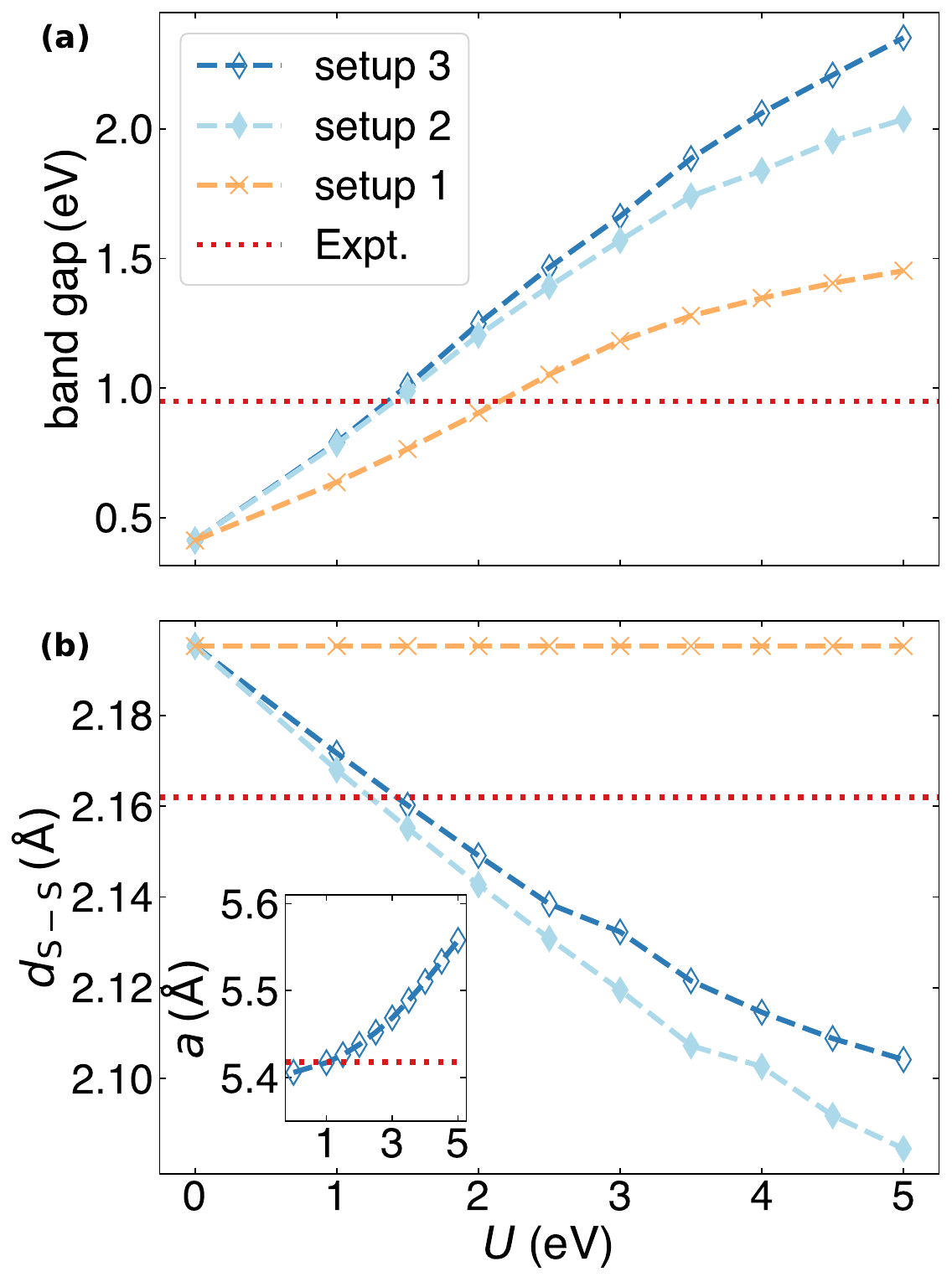}
                \caption{\textbf{(a)} calculated band gaps and \textbf{(b)} \ce{S-S} bond lengths computed with PBE$+U$ using three setups: at fixed cell and ionic positions (setup 1); at constant cell volume relaxing the ionic positions (setup 2); and relaxing both the ionic positions and the cell volume (setup 3). The inset in the bottom panel shows the dependence of the lattice parameter on $U$ during volume relaxations. Red dashed horizontal lines correspond to the experimental values\cite{Zuniga_pyrite_2019,ENNAOUI_1984}.}
                \label{fig:gap_vs_lattice}
            \end{figure}

            Figure~\ref{fig:gap_vs_lattice} illustrates that even when no ionic relaxation is considered (setup 1), the band gap slowly but steadily expands with increasing $U$ values. The shifts in band eigenvalues indicate that this expansion primarily stems from a downshift of the $t_{g}$ orbitals' KS potential in the valence region (see Table SI 1). 
            On the other hand, the CBM remains largely unaffected by the Hubbard correction with the corresponding eigenvalues showing minimal changes. 
            Consistent with previous studies, we find that the experimental band gap is accurately reproduced at $U\approx 2.0\,$eV, but is overestimated by approximately 53\% at $U=5.0\,$eV. This overestimation of the band gap at such moderate values of $U$ is surprising since the CBM, consisting of S-$3p_z$ states, is not included in the Hubbard manifold, and its local deviation from PWL remains uncorrected.
            Upon relaxing the ionic positions while keeping the cell volume constant (setup 2), the band gap shows a more sensitive response to higher values of $U$. The experimental band gap is already achieved at $U\approx 1.5\,$eV and is overestimated by 115\% at $U=5.0\,$eV. This trend becomes even more pronounced when relaxation of the cell vectors is enabled (setup 3), resulting in a band gap overestimation of 147\% at $U=5.0\,$eV.
            As shown in Figure~\ref{fig:gap_vs_lattice}(b), the additional expansion following ionic relaxations is correlated with a significant contraction of the \ce{S-S} bonds. This aligns well with the findings of \citeauthor{eyert_electronic_1998}, who pointed out that the \ce{S-S} bond length governs the band gap, as it controls the dispersion of the lowest conduction band around the $\Gamma$ point~\cite{eyert_electronic_1998}.
            Remarkably, in variable-cell calculations (setup 3) the \ce{S-S} bonds continue to contract with increasing values of $U$, despite the simultaneous growth of the lattice parameter (inset of Figure~\ref{fig:gap_vs_lattice}(b)).
            The differences observed between relaxed and fixed structure calculations reveal that shell-averaged $U$ corrections modify the band gap through two distinct mechanisms: first, through the expected downshift of the valence band eigenvalues, and second by causing pronounced contractions of the \ce{S-S} bonds.
            
            We now shift our focus from conventional equilibrium properties to investigate the underlying reasons behind the substantial influence of shell-averaged $U$ corrections on these bonds, which can be elucidated by examining the occupations of the individual orbitals.
            Table~\ref{Tab:pyrite_eigvals} presents the eigenvalues of the Fe-$d$ occupation matrix and the corresponding Hubbard energies $E_U$ for an illustrative case with $U=3.0\,$eV, both prior to and after complete structural relaxation.
            As the eigenvalues of the $t_{g}$ orbitals ($d_{xy}$, $d_{xz}$, $d_{yz}$) approach idempotency, this manifold is responsible for the smaller share of the overall Hubbard energy, approximately 33\%.
            In contrast, the $e_g$ orbitals ($d_{x^2-y^2}$ and $d_{z^2}$) contribute to nearly 66\% of the total Hubbard energy due to possessing occupation eigenvalues far from either 0 or 1.
            Structural relaxation induces both intra-shell and inter-shell charge transfers that allow for a 6\% reduction of the total Hubbard energy.
            At the intra-shell level, the occupation of $t_g$ orbitals grows at the expense of $e_g$. 
            Even more significantly, the overall $d$-occupancy slightly drops from 6.996 to 6.970, indicating a migration of some $e_g$ electrons into adjacent S-$p$ orbitals.
            This transfer of charge effectively enhances the \ce{S-S} bond order and leads to the observed contraction of $d_{\mathrm{S-S}}$.
            Thus, the unexpectedly potent influence of the value of $U$ on the band gap is rooted in the correction of the $e_g$ manifold.
            \begin{table}[ht]
                \caption{Occupation eigenvalues $\lambda$ and corresponding Hubbard energies $E_U$ obtained for the Fe-$d$ orbitals in \ce{FeS2} using PBE$+U$ with $U=3.0\,$eV. Results are shown for the fixed PBE structure (setup 1, ${a=5.406}$\,\AA~ and a \ce{S-S} distance of 2.195\,\AA, top) and after structural relaxation (setup 3, $a=5.469$\,\AA~ and a \ce{S-S} distance of 2.132\,\AA, bottom). The last column shows the total occupations ($2\sum \lambda$) and Hubbard energies ($2\sum E_U$), which correspond to twice the sum (due to spin degeneracy) of the individual contributions.}
                \begin{tabular}{@{}clllllll@{}}
                    \toprule
                    & Orbital &  $d_{x^2-y^2} $ & $d_{z^2} $ & $d_{xy}$ & $d_{xz}$ & $d_{yz}$ & $2\Sigma$ \\ \midrule
                    setup 1  & $\lambda$   & 0.383 & 0.383 & 0.901 & 0.901 & 0.930 & 6.996     \\
                    & $E_U$ (meV)          & 354   & 354   & 134   & 134   & 98    & 2148      \\ \midrule
                    setup 3  & $\lambda$ & 0.359 & 0.359 & 0.915 & 0.915 & 0.937 & 6.970     \\
                    & $E_U$ (meV) & 343  & 343   & 117   & 117   & 89    & 2018      \\ \bottomrule 
                \end{tabular}
                \label{Tab:pyrite_eigvals}
            \end{table}
        
            We recall that the eigenvalues listed in Table~\ref{Tab:pyrite_eigvals} are not universal quantities but the result of a projection of the KS wave functions onto atom-centered OAO.
            The accuracy of on-site occupations provided by atomic-like projectors relies on the similarity between a given orbital's shape in the system under inspection and that in a free atom, since the general shape and extension of projector AOs is typically determined in free atom calculations.
            However, one-center projections using atomic orbitals struggle to account for strong orbital hybridization, where electrons localize `off-site' between the bonded atoms.
            In the case of \ce{FeS2}, the $t_{g}$ manifold is only marginally bonding, while the formally unoccupied $e_g$ orbitals form $\sigma$ MOs with neighboring S-$3p_z$ orbitals. Consequently, it is misleading to interpret the eigenvalues corresponding to $d_{x^2-y^2}$ and $d_{z^2}$ as indicative of actual on-site orbital occupancies in the context of Hubbard $U$ corrections.
            Moreover, their numerical values significantly hinge on computational factors like the chosen pseudopotentials and the specific charge state for which they were parameterized~\cite{eyert_electronic_1998,kulik_self-consistent_2008}.

            Within the setup of this study, the $U$ correction effectively penalizes the hybridization between \ce{Fe} and \ce{S}; however, the use of different Hubbard projector functions cause the opposite effect, for example, if NAO projectors yield $e_g$ eigenvalues larger than $0.5$. In this case, the Hubbard correction would draw electrons into the $e_g$ manifold rather than expelling them (see Eq. \eqref{eq:hubbard-potential-u}).
            
        \subsubsection{Absence of intra-shell screening in shell-averaged LR-cDFT}
        \label{subsec:lr-cdft-failure}
            Having investigated how the shell-averaged $U$ parameter affects equilibrium observables, our focus now shifts to understanding the implications when this parameter is determined from first principles.
            Applying DFPT to the PBE ground state of \ce{FeS2} yields a $U$ value of $7.37\,$eV. After undergoing four iterations within the self-consistency loop detailed in Section~\ref{subsec:lr-cdft}, $U$ converges to $6.47\,$eV.
            Unfortunately, neither of these parameters can reasonably reproduce the experimental characteristics of pyrite. In fact, both values result in the stabilization of a spurious ferromagnetic ground state ($2\,\mu_B/$cell) in unrestricted open-shell calculations. For comparison, all of the following $U$ parameters and observables are reported for the nonmagnetic ground state that was enforced by fixing the total magnetization to 0.0$\,\mu_{\mathrm{B}}$.
            \begin{figure}[h]
                \includegraphics[width=3.33in]{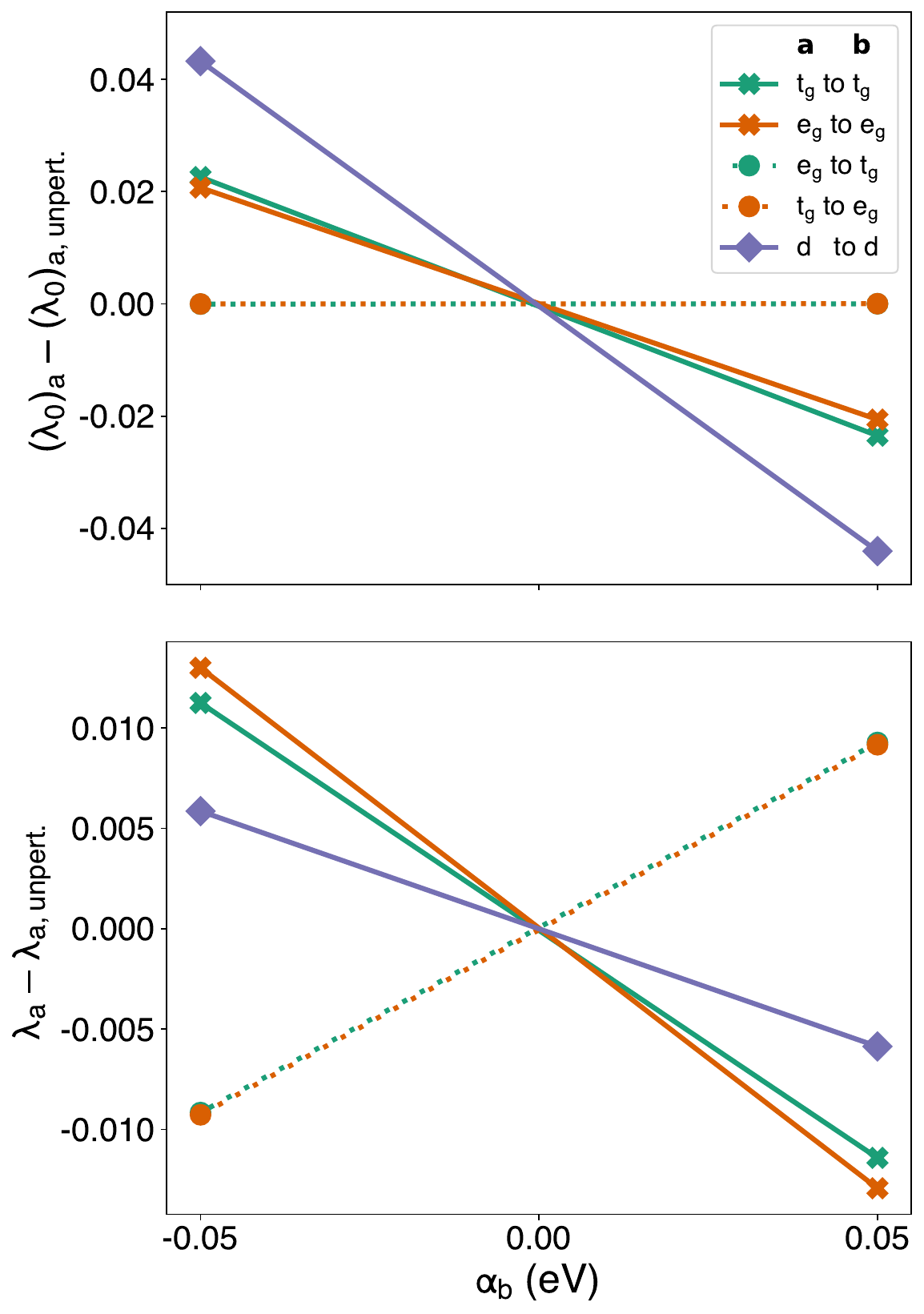}
                \caption{Unscreened/non-interacting (top) and screened/interacting (bottom) responses of the occupations $\lambda$ of  manifolds ``$\mathbf{a}$'' to perturbations $\alpha$ applied to manifolds ``$\mathbf{b}$'' in \ce{FeS2}, obtained through the orbital-resolved LR-cDFT approach. The starting point is the fully relaxed PBE ground state.}
                \label{fig:response}
            \end{figure}

            The reason for this significant overestimation of the shell-averaged Hubbard $U$ parameter can be understood from Figure~\ref{fig:response}, which shows the response of orbital occupancies to perturbations for both the entire $3d$-shell and for its irreducible representations $t_g$ and $e_g$. 
            This information cannot be extracted from shell-averaged LR-cDFT, but can be recovered by adopting the orbital-resolved approach.
            The substantial opposing responses of $e_g$ occupations to perturbations of $t_g$ and vice versa (depicted by dashed lines in the bottom panel) suggest the presence of a robust intra-shell $t_{g} \leftrightarrow e_g$ screening channel.
            Inserting the values of $\chi_0$ and $\chi$ (represented by the line slopes in Figure \ref{fig:response}) into Eq. \eqref{eq:pickett}, one obtains
            \begin{eqnarray}
                U & = & \left[\sum
                \begin{pmatrix}
                -0.4608 &  0.0001    \\
                0.0001 & -0.4131
                \end{pmatrix}\right]^{-1}
                 - \left[\sum
                \begin{pmatrix}
                -0.2266	&  0.1843    \\
                0.1843 & -0.2592
                \end{pmatrix}\right]^{-1} \nonumber \\ [10pt]
                & = &
                (-0.8737)^{-1} - (-0.1172)^{-1} =
                -1.144 - (-8.532) = 7.388\,\mathrm{eV},
                \label{eq:u-pyrite}
            \end{eqnarray}
            where the first and second matrix represent the unscreened and the screened responses, respectively. Note the almost perfect agreement between this LR-cDFT Hubbard parameter ($U = 7.39\,$eV) and the one derived from DFPT ($U = 7.37\,$eV); the minor discrepancies are due to numerical noise. 
            The individual matrix elements of Eq.~\eqref{eq:u-pyrite} reveal that the apparent overestimation of the shell-averaged Hubbard parameter primarily stems from the suppression of intra-shell screening that occurs due to the simultaneous perturbation of the $t_{g}$ and the $e_g$ manifolds.
            While the individual screened responses of the $t_{g}$ and $e_g$ orbitals are substantial ($\chi_{t_g,t_g}= -0.2266\,\mathrm{eV}^{-1}$ and $\chi_{e_g,e_g}= -0.2592\,\mathrm{eV}^{-1}$), the overall response of the $d$ shell is massively reduced by the off-diagonal elements of $\chi$ (both $0.1843\,\mathrm{eV}^{-1}$). 
            Inversion of this \textit{apparently} small screened response of the $d$ shell ($-0.1172\,\mathrm{eV}^{-1}$) yields $\chi^{-1}=-8.532\,$eV, which is what ultimately fuels the overestimation of the shell-averaged Hubbard parameter.

            It should be noted that this shell-averaged $U$ parameter also encompasses, to a certain extent, the response of the S-$3p$ orbitals, because the $e_g$ eigenstates likely contain substantial S-$3p$ contributions.
            With these insights in mind, it becomes pertinent to investigate whether the orbital-resolved approach to Hubbard $U$ can offer improvements over the unsatisfactory performance of shell-averaged PBE$+U$.
            
        \subsubsection{Application of the orbital-resolved \texorpdfstring{$U$}{U}}
        \label{subsec:um-to-pyrite}
            Employing the orbital-resolved LR-cDFT methodology detailed in Section~\ref{subsec:lr-cdft}, we proceed to compute individual Hubbard parameters for the $t_g$ and $e_g$ orbitals of \ce{FeS2}. Prior to inverting the response matrix, all off-diagonal intra-shell matrix elements are set to zero:
            \begin{equation}
                (\chi)^{II}_{i\neq j} = 0 \textnormal{ , and  } (\chi_0)^{II}_{i\neq j} = 0 .
            \end{equation}
            This ensures that the resulting orbital-resolved Hubbard parameters incorporate the effect of $t_{g} \leftrightarrow e_g$ screening. The parameters $U_{t_{g}}$ and $U_{e_{g}}$ converge rapidly, reaching their self-consistent values of $3.29\,$eV and $2.16\,$eV, respectively, within just three iterations of self-consistency. The final response matrices read 
            \begin{eqnarray}
                \begin{pmatrix}
                U_{t_{g}} & 0 \\
                    0      & U_{e_{g}}
                \end{pmatrix} 
                & = & \left[
                \begin{pmatrix}
                -0.3516 &  0    \\
                     0  & -0.1948
                \end{pmatrix}\right]^{-1}
                 - \left[
                \begin{pmatrix}
                -0.1998	&  0    \\
                  0     & -0.1187
                \end{pmatrix}\right]^{-1} \nonumber \\ [10pt]
                & = &
                \begin{pmatrix}
                3.29 & 0 \\
                  0  & 2.16
                \end{pmatrix} .
            \end{eqnarray}
            Following the approach of \citeauthor{solovyev_t_1996}, one may also exclusively target the $t_g$ manifold and not correct the $e_g$ states at all. This choice is driven by the expectation that interactions of genuine \textit{on-site} character should manifest within the occupied $t_{g}$ orbitals rather than in the hybridized and formally empty $e_g$ manifold. Moreover, as previously stated, it is unlikely that one-center atomic orbital projectors are suited to provide meaningful on-site occupation numbers for the $e_g$ orbitals. 
            With $U_{e_{g}}$ corrections absent, the converged value of $U_{t_{g}}$ slightly decreases to $3.01\,$eV.
            \begin{table}[h]
                \caption{Comparison of the equilibrium lattice parameter $a$, the \ce{S-S} bond length $d_{\mathrm{S-S}}$ and the band gap $E_{\mathrm{g}}$ of pyrite derived from Hubbard-corrected PBE calculations targeting different manifolds. On-site Hubbard $U_1$ and $U_2$ refer to various parametrizations of the PBE+$U$ approach, while $V$ is the inter-site Hubbard parameter of the PBE+$U$+$V$ approach.}
                \begin{tabular}{@{}lccccccc@{}}
                    \toprule
                    & PBE & $+U$ & ${+U \! + \! V}$ &$+U^{\mathrm{emp}}$ & $+U_{t_g}+U_{e_g} $ & $+U_{t_g}$ & {Expt.} \\ \midrule
                    $U_1\,$(eV)                 & ---       & 6.47     & 6.73  & 1.50  & 3.29 & 3.01   & ---    \\
                    $U_2$ or $V\,$(eV)             & ---       & ---      & 0.62  & ---   & 2.16 & ---    & ---     \\ \midrule
                    $a\,$(\AA)                  & 5.41      & 5.63     & 5.51  & 5.43  & 5.46 & 5.44   & 5.42\cite{Zuniga_pyrite_2019}    \\
                    $d_{\mathrm{S-S}}\,$(\AA)   & 2.20      & 2.10     & 2.12  & 2.16  & 2.13 & 2.14   & 2.16\cite{Zuniga_pyrite_2019}    \\    
                    $E_{\mathrm{g}}\,$(eV)      & 0.41      & 2.63     & 2.06  & 1.00  & 1.70 & 1.45   & 0.95\cite{ENNAOUI_1984}    \\ \bottomrule
                \end{tabular}
                \label{tab:pyrite_results}
            \end{table}
            Having established the necessary parameters, we now focus on their impact in practical calculations.
            Table~\ref{tab:pyrite_results} shows that the orbital-resolved approach consistently outperforms the shell-averaged approach, except when $U$ is tuned empirically ($U^{\mathrm{emp}}$). 
            When utilizing only $U_{t_{g}}$, both the lattice parameter and the \ce{S-S} bond length exhibit deviations of less than 1\% from the experimental values. With orbital-resolved corrections applied to ${t_{g}}$ and $e_g$, this agreement with the experimental data is slightly worsened but still surpasses that of shell-averaged PBE$+U$, where the lattice parameter deviates by 4\%.
            Furthermore, the significant overestimation of the band gap amounting to 177\% in the case of shell-averaged PBE$+U$ is substantially mitigated to 53\% when switching to $U_{t_{g}}$. 
            The DFT${+U \! + \! V}$ approach yields noticeable improvements compared to the traditional $+U$ approach; however, it appears that the inter-site $V$ term lacks the strength to fully restore the hybridization between \ce{Fe} and \ce{S} that is suppressed by the shell-averaged on-site term. Therefore, the estimation of the lattice parameters and the band gap still remains notably worse than for the orbital-resolved PBE$+U$ approach.
    
            These results underscore the critical importance of the choice of the Hubbard manifold in DFT$+U$ calculations for the \ce{FeS2} polymorph pyrite. Under the shell-averaged approximation, the ground-state properties are extremely sensitive to the numerical value of $U$. This sensitivity primarily arises due to the correction of the hybridized $e_g$ part of the $d$ shell.  
            Moreover, the shell-averaged Hubbard parameter derived from LR-cDFT lacks full screening, as the intra-shell screening channel ($t_{g} \leftrightarrow e_g$) is deactivated during the concurrent perturbation of both manifolds.
            On the other hand, the orbital-resolved approach accounts for intra-shell screening during perturbative calculations, resulting in significantly smaller Hubbard parameters. We have shown that correcting the localized Fe-$t_{g}$ manifold alone suffices to obtain structural properties in very good agreement with experimental data, all without requiring empirical adjustments to the parameters. A viable option for further improvements involves supplementing the orbital-resolved $U$ corrections with inter-site $V$ terms, as discussed in Section~\ref{sec:final_remarks}. Additionally, the PWL condition should also be enforced for the \ce{S}-$3p_z$ orbital composing the CBM. Given the strong $sp^3$ hybridization of \ce{S}, this would, however, require an adaption of the atomic-like projectors used in this study.

    \subsection{Fe(II) molecular complexes}
        \label{subsec:fe2-complexes}
        \subsubsection{Adiabatic spin energy differences}
        A crucial quantity for many TM compounds is the total energy difference between their HS and LS states, denoted as:
        \begin{equation}
            \Delta E_{H-L} = E^{\mathrm{HS}}_{tot } - E^{\mathrm{LS}}_{tot} \textnormal{  .}
        \end{equation}
        Based on calculations of six \ce{Fe{(II)}}-hexacomplexes, \citeauthor{mariano_biased_2020} showed that shell-averaged Hubbard $U$ corrections to PBE and LDA introduce an unphysical bias against LS configurations, primarily due to substantial discrepancies in the Hubbard energies of LS and HS complexes~\cite{mariano_biased_2020}.
        This arises from the electron occupation patterns: LS-\ce{Fe{(II)}} complexes exhibit a  $t_{2g}^6 / e_g^0$ electron configuration, while HS complexes a $t_{2g}^4 / e_g^2$ configuration.
        In octahedral coordination environments, \textit{unoccupied} $e_g$ orbitals display substantial $\sigma$ overlap with neighboring ligand orbitals.
        Consequently, in LS-\ce{Fe{(II)}} all four $e_g$ spin-orbitals can adopt fractional occupation eigenvalues, whereas in HS-\ce{Fe{(II)}} this only applies to two spin-orbitals, as the other two are fully occupied.
        Considering that the Hubbard energy functional (Eq.~\eqref{eq:hubbardU-1-Eu}) induces corrections of up to $U/2\,$eV for fractional occupations, but of $0\,$eV for idempotent occupations, Hubbard energies of LS-\ce{Fe{(II)}} tend to be larger than those of HS-\ce{Fe{(II)}}.
        Furthermore, the LR-cDFT approach was shown to generate larger shell-averaged $U$ parameters for LS-\ce{Fe{(II)}} compared to HS-\ce{Fe{(II)}}~\cite{mariano_biased_2020}, thus enhancing this disparity.

        In the following, we investigate whether the orbital-resolved DFT$+U$ approach can rectify the unphysical bias against LS states and deliver accurate adiabatic spin energy differences.
        For this purpose, we compute $\Delta E_{H-L}$ for all six octahedrally coordinated \ce{Fe{(II)}}-hexacomplexes studied in Ref. \citenum{mariano_biased_2020} and benchmark the results against coupled-cluster corrected CASPT2 (CASPT2/CC)\cite{phung_toward_2018} values, which are also taken from Ref. \citenum{mariano_biased_2020}.
        We compare the performance of two distinct Hubbard manifolds.
        In \textbf{manifold I}, Hubbard corrections are applied either only to $e_{g}$ states (\ce{[Fe(H2O)_6]^{2+}} and \ce{[Fe(NH3)_6]^{2+}}) or to only $t_{2g}$ states (\ce{[Fe(NCH)_6]^{2+}}, \ce{[Fe(PH3)_6]^{2+}},  \ce{[Fe(CO)_6]^{2+}}, \ce{[Fe(CNH)_6]^{2+}}). While it would be desirable to obtain $t_{2g}$-specific Hubbard parameters for the former two compounds, this proves elusive due to the fact that the $t_{2g}$ orbitals are fully occupied in the numerical sense ($\lambda \geq 0.997$) and thus respond non-linearly to perturbations~\cite{kulik_systematic_2010}.
        For the same reason, no $U_{t_{2g}}^{\mathrm{HS}}$ parameters could be computed for the HS complexes either. Therefore, all Hubbard corrections to $t_{2g}$ states utilize the $U_{t_{2g}}^{\mathrm{LS}}$ value.
        In \textbf{manifold II}, both $t_{2g}$ and $e_g$ orbitals are corrected simultaneously.
        When comparing the present work with Ref.~\citenum{mariano_biased_2020}, it is important to note that we use OAO projector functions and self-consistent procedures for geometry relaxations and determination of Hubbard terms.
        A comprehensive list of all complexes, the Hubbard manifolds considered and the corresponding orbital-resolved and shell-averaged $U$ parameters is provided in Table~\ref{tab:manifolds}.
        \begin{table}[t]
            \caption{PBE+$U$ manifolds and their calculated on-site parameters in eV. For $t_{2g}$ the same parameter was applied to LS and HS complexes. The complexes are sorted in ascending order according to the ligand's field strength.}
            \begin{tabular}{@{}lccccccc@{}}
                \hline
                Complex & manifold I & manifold II   & $U_{t_{2g}}^{\mathrm{LS}}$ & $U^{\mathrm{LS}}$ & $U^{\mathrm{HS}}_{e_g}$ & $U^{\mathrm{LS}}$ & $U^{\mathrm{HS}}$ \\ \midrule
                \ce{[Fe(H2O)_6]^{2+}} & $e_g$   & ---              & ---  & 4.28 & 2.69 & 5.71 & 4.13     \\
                \ce{[Fe(NH3)_6]^{2+}} & $e_g$   & ---              & ---  & 3.64 & 3.07 & 6.14 & 4.43     \\
                \ce{[Fe(NCH)_6]^{2+}} & $t_{2g}$& $t_{2g}$+$e_g$   & 7.05 & 2.12 & 1.87 & 6.76 & 5.53     \\
                \ce{[Fe(PH3)_6]^{2+}} & $t_{2g}$& $t_{2g}$+$e_g$   & 4.13 & 2.29 & 2.12 & 6.88 & 4.78     \\
                \ce{[Fe(CO)_6]^{2+}}  & $t_{2g}$& $t_{2g}$+$e_g$   & 4.63 & 1.92 & 1.89 & 7.16 & 5.43     \\
                \ce{[Fe(CNH)_6]^{2+}} & $t_{2g}$& $t_{2g}$+$e_g$   & 4.64 & 2.13 & 1.72 & 7.43 & 5.79     \\ 
                \hline
            \end{tabular}
            \label{tab:manifolds}
        \end{table} 

\begin{figure}[t]
            \includegraphics[width=3.33in]{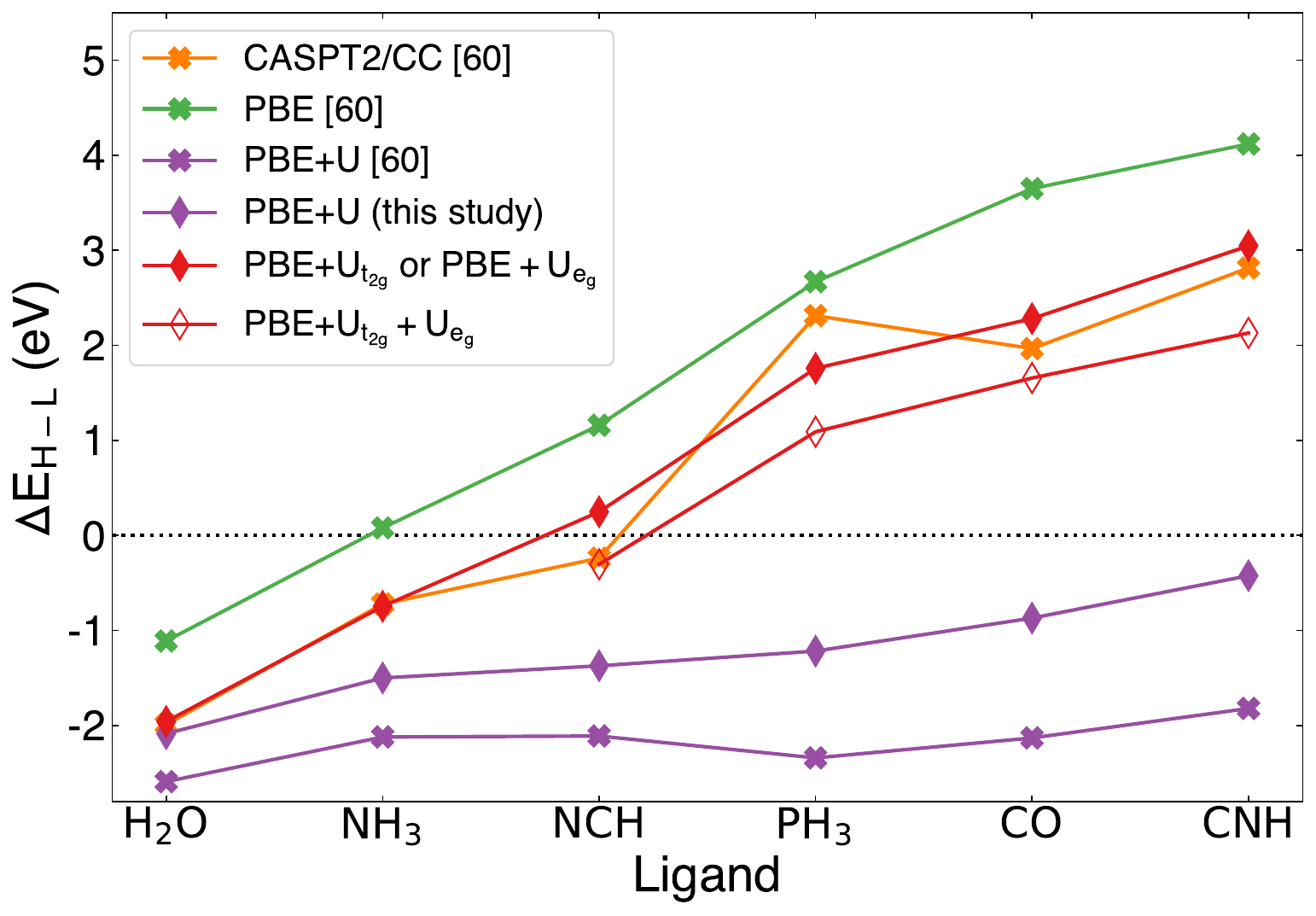}
            \caption{Adiabatic spin energies calculated for six Fe(II)-hexacomplexes using different DFT corrections and CASPT2/CC as the reference method. Data for PBE, PBE+$U$, and CASPT2/CC was taken from the SI of Ref.~\citenum{mariano_biased_2020}.}
            \label{fig:spin-energies}
        \end{figure}

        The resulting adiabatic spin energy differences are visualized in Figure~\ref{fig:spin-energies}. Additionally, Table~\ref{tab:mae} lists the mean absolute error (MAE) of the different corrections against the CASPT2/CC\cite{phung_toward_2018} reference values.
        In line with the findings of \citeauthor{mariano_biased_2020}, shell-averaged PBE$+U$ drastically overstabilizes HS configurations and fails to reproduce the trend of increasing $\Delta E_{H-L}$. Conversely, uncorrected PBE displays its known bias towards LS configurations.
        The introduction of orbital-resolved $U$ corrections substantially enhances the predictive accuracy across all complexes analyzed, outperforming both PBE and the shell-averaged PBE$+U$ by a large margin.
        Correcting either $t_{2g}$ or $e_g$ (manifold I) closely aligns adiabatic spin energy differences with the CASPT2/CC values, resulting in the lowest MAE of $0.27\,$eV. 
        The simultaneous correction of $t_{2g}$ and $e_g$ (manifold II) yields a slightly higher MAE of $0.57\,$eV, with larger deviation particularly for \ce{[Fe(PH3)6]^{2+}} and \ce{[Fe(CNH)6]^{2+}}, but a better agreement for \ce{[Fe(NCH)_6]^{2+}}. 
        In general, $U_{e_g}$ corrections are smaller and less significant for the overall quantitative agreement than $U_{t_{2g}}$ corrections, since the $t_{2g}$ orbitals host more -- and more localized -- electrons.
        The good overall agreement (qualitative \textit{and} quantitative) of orbital-resolved PBE$+U$ with the CASPT2/CC data is surprising since the latter is a multiconfigurational approach, while the former can only improve the accuracy of a single reference state.
        
        \begin{table}[t]
            \caption{Mean absolute error (MAE) of $\Delta E_{H-L}$ for various Hubbard correction manifolds against the CASPT2/CC reference value from Ref.~\citenum{mariano_biased_2020}.}
            \begin{tabular}{@{}llllll@{}}
            \toprule
                     &  PBE\cite{mariano_biased_2020}  & +$U$   &  +$U_{t_{2g}}$ or +$U_{e_g}$  & +$U_{t_{2g}}$+$U_{e_g}$ \\ \midrule
            MAE (eV) & \multicolumn{1}{c}{1.07} & \multicolumn{1}{c}{1.94} & \multicolumn{1}{c}{0.27} & \multicolumn{1}{c}{0.57} \\ \bottomrule
            \end{tabular}
            \label{tab:mae}
        \end{table}    
        The good performance of the orbital-resolved $U$ compared to the shell-averaged approach stems from two main factors. First, excluding the $e_g$ orbitals from the Hubbard manifold of strong-field complexes (manifold I) eliminates the primary cause of spuriously large Hubbard energies. Again, the exclusion of $e_g$ orbitals can be justified with the inadequate representation of their strongly hybridized character by atomic orbital projectors.
        Second, even when the $e_g$ orbitals are included (manifold II), independently computed $U_{e_g}$ parameters are significantly smaller than   $U_{t_{2g}}$ or shell-averaged $U$ parameters owing to the explicit incorporation of intra-shell screening. For instance, in \ce{[Fe(CO)_6]^{2+}} $U_{e_g}^{\mathrm{LS}}=1.92\,$eV compared to $U^{\mathrm{LS}}=7.16\,$eV.
        Hence, the orbital-resolved approach effectively diminishes the impact of non-ideal Hubbard projectors. 

        \subsubsection{Piece-wise linearity of the total energy}
        As mentioned earlier, the use of on-site Hubbard $U$ corrections in (semi-)local DFT is motivated by the mitigation of SIE, which have been linked to the spurious global deviation from PWL of the total energy with respect to fractional addition or removal of electronic charge $q$ to the entire system \cite{kulikDensityFunctionalTheory2006,zhao_global_2016,bajaj_communication_2017}.
        However, a study by \citeauthor{zhao_global_2016} indicates that (shell-averaged) $U$ corrections to TM \textit{d} shells might be unfit for this purpose for a wide range of TM/ligand combinations when the Hubbard $U$ parameters are derived from LR-cDFT\cite{zhao_global_2016}.
        
        Thus, to directly benchmark the performance of orbital-resolved DFT+$U$, we explicitly determine the global curvature following the approach presented in Ref.~\citenum{zhao_global_2016}. 
        For our analysis, we chose the strongest ligand complex \ce{[Fe(CNH)_6]^{q+}} and perform several fixed-charge calculations where the total charge is varied between $q=2$ and $q=3$ in increments of $0.1$e$^{-}$.
        \begin{figure}[t]
            \includegraphics[width=6in]{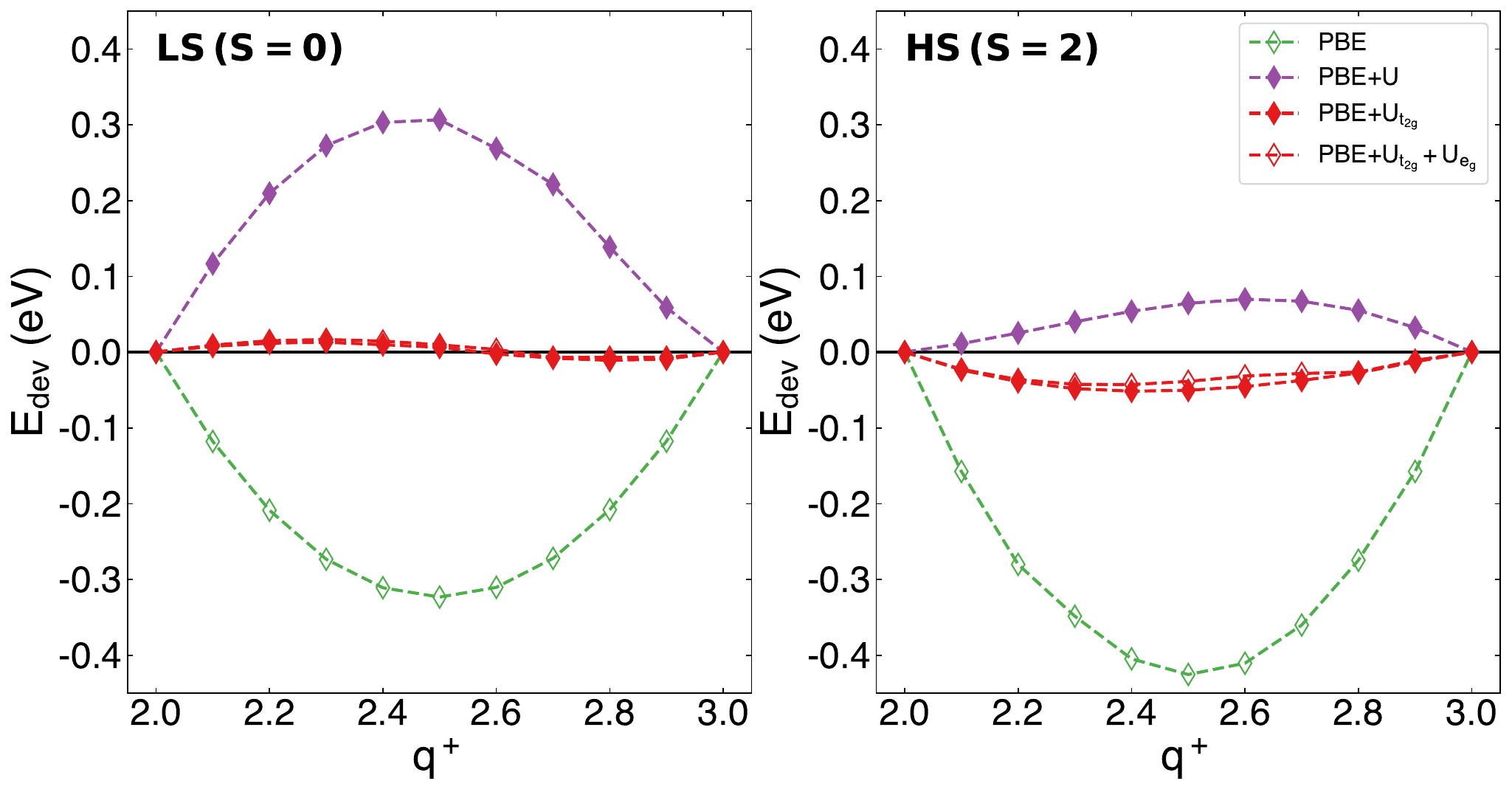}
            \caption{Deviation from piecewise linearity of the total energy upon fractional addition/removal of charge of \ce{[Fe(CNH)6]^{(q+)}} in low-spin (left) and high-spin (right) configurations. Note that the data points of PBE$+U_{t_{2g}}$ and PBE$+U_{t_{2g}}+U_{e_g}$ overlap in the LS case.}
            \label{fig:piecewise-linearity}
        \end{figure}
        A useful metric to assess a functional's deviation from PWL is given by $E_\mathrm{dev}$, which is computed by subtracting the DFT total energy of a fractional charge calculation from the linear interpolation between the energies of the integer-charge endpoints $E_{tot}(q=2)$ and $E_{tot}(q=3)$.
        Note that this explicit approach differs from the method employed by \citeauthor{mariano_biased_2020}, who approximate $E_\mathrm{dev}$ using a cubic interpolation parameterized by the energy difference $\Delta E_q = E_{tot}(q=2)- E_{tot}(q=3)$ and the HOMO and LUMO eigenvalues of the integer-charge system \cite{mariano_biased_2020}.
        
        Figure~\ref{fig:piecewise-linearity} shows that while shell-averaged Hubbard $U$ corrections markedly reduce the deviation from PWL of HS \ce{[Fe(CNH)_6]^{q+}}, the global curvature of the LS compound is not eased.
        In fact, in the LS case the typical convex-shaped FCE displayed by bare PBE is transformed into a concave one of similar magnitude. 
        In contrast, the orbital-resolved corrections reduce the deviation from PWL by an order of magnitude for both the LS and the HS configurations.
        It is worth noting that the remaining curvature barely differs between manifold I (+$U_{t_{2g}}$ only) and manifold II (+$U_{t_{2g}}$ \! + \! $U_{e_g}$).
        Again, this observation is likely related to the low numerical value of the $U_{e_g}$ parameters, which further supports the previous assumption that the primary contributor to the FCE is the rather localized $t_{2g}$ manifold. 
        For the same reason, applying $U$ corrections to ligand orbitals like \ce{C}-$p$ is unlikely to significantly affect the FCE.

    \subsection{The importance of correcting ligand states in \texorpdfstring{\ce{\beta-MnO2}}{Beta-MnO2}}
        Unlike in TM sulfides, partially filled \textit{d} shells of TM oxides are a frequent target for on-site corrections (see the discussion e.g. in Ref.~\citenum{Gebreyesus:2023}). While Mott-Hubbard insulators such as the prototypical w\"ustite (\ce{FeO}) present an energy gap separating different $d$ states, so-called charge-transfer insulators, often involving highly oxidized and/or heavy TM species, exhibit \ce{O}-$2p \rightarrow d$ band gaps~\cite{zaanen_band_1985}.
        The significant fraction of electrons localized on \ce{O}-$2p$ orbitals adds complexity to the definition of the Hubbard manifold for DFT$+U$ calculations, as studied below for the case of \ce{\beta-MnO2}.        
        
        Crystallizing in the rutile structure (space group P$4_2$/mnm), \ce{\beta-MnO2} displays a complex helical (screw-type spiral) antiferromagnetic (AFM) ordering below $T_N = 92\,$K \cite{sato_transport_2000}. As done here, this ordering can be approximated by a collinear arrangement of the \ce{Mn^{4+}} ions along [001], referred to as A1-AFM\cite{mahajan_pivotal_2022}.
        Considerable uncertainty surrounds the electronic properties of \ce{\beta-MnO2}. \citeauthor{sato_transport_2000} measured a large electrical resistivity at $0.3\,$K, suggesting insulating behavior \cite{sato_transport_2000}. However, the magnitude of the band gap has not been yet determined with universally accepted accuracy.
        Mid-twentieth-century works reported narrow band gaps around $0.26-0.28\,$eV~\cite{chevillot_influence_1959,druilhe_semiconducteurs_1966}, while more recent optical absorption studies reported much larger values ranging from $1.5\,$eV (nanocacti and nanorods)~\cite{kumar_morphology_2017} to $2.0\,$eV for \ce{\beta-MnO2} nanostructures grown on fluorine-doped tin oxide~\cite{bigiani_dual_2020}.
        These results align with a reported hybrid functional DFT (PBE0) value of $1.5\,$eV, but not with a prediction of the closely related HSE03 functional ($0.6\,$eV)~\cite{franchini_ground-state_2007}.
        Previous works employing the shell-averaged DFT+$U$ approach suggest that a band gap is not opened unless the Mn-$d$ states are corrected using $U$ values larger than 6\,eV\cite{franchini_ground-state_2007,mahajan_importance_2021}.
        In general, the calculated properties are sensitive to the choice of projector functions. For instance, NAO projectors stabilize a ferromagnetic ordering, whereas OAO projectors favor the expected A1-AFM ordering~\cite{mahajan_importance_2021}.
        Only by adding inter-site $V$ between Mn-$d$ and O-$p$ states~\cite{mahajan_importance_2021} or Hund's $J$ corrections~\cite{lim_improved_2016}, small gaps ranging from $0.25$ to $0.32\,$eV emerge.
        A band gap of $0.8\,$eV was also obtained~\cite{tompsett_importance_2012} with the anisotropic DFT+$U$+$J$ approach of \citeauthor{czyzyk_local-density_1994}. In this method, shell-averaged Hubbard $U$ and $J$ corrections are augmented with orbital-resolved $U_{mm'}$ and $J_{mm'}$ matrices parameterized from summation relations involving Slater integrals of atomic Hartree-Fock calculations~\cite{czyzyk_local-density_1994}. 
        This approach thus differs from the here-presented scheme, which does not rely on model systems like the free atom and includes intra-manifold screening effects, such as $t_{2g} \leftrightarrow e_g$.          
        
        The wide range of band gap values reported in both the theoretical and experimental literature underscores the pressing need for a more profound understanding of the insulating behavior of \ce{\beta-MnO2}. 
        Considering the charge-transfer nature of the band gap and the coexistence of relatively localized and strongly hybridized states in both the \ce{Mn}-$d$ shell and the \ce{O}-$p$ shell, orbital-resolved DFT+$U$ calculations offer a promising starting point.
        In the interest of comparability, our calculations are carried out using a setup equivalent to that of Ref. \citenum{mahajan_importance_2021}, imposing the A1-AFM ordering in a $2\times2\times2$ supercell containing 48 atoms and considering once again different Hubbard manifolds.

        \begin{table*}[!ht]
        \caption{Comparison of equilibrium properties of A1-AFM \ce{\beta-MnO2} obtained from PBEsol without and with various Hubbard corrections. $U_1$ and $U_2$ refer to various parameterizations of the PBEsol+$U$ approach, while $V$ is the inter-site Hubbard parameter of the PBEsol+$U$+$V$ approach. $a$ and $c$ are the lattice parameters, $V_0$ is the unit cell volume, and $E_g$ is the band gap value. All the presented results (including those from Ref.~\cite{mahajan_importance_2021}) were obtained using OAO projectors.
        }
        \label{tab:mno2-results}
        \begin{tabular}{@{}lcccccc@{}}
        \toprule
        & PBEsol & +$U^{\mathrm{Mn}}_{t_{2g}}$ & \parbox{2.3cm}{\centering $+U^{\mathrm{Mn}}_{t_{2g}}+U^{\mathrm{O}}_{p_z}$} & \parbox{1.7cm}{$+U^{\mathrm{Mn}}$ \\ (Ref. \citenum{mahajan_importance_2021})} & \parbox{1.7cm}{${+U^{\mathrm{Mn}} \! + \! V^{\mathrm{Mn-O}}}$ \\ (Ref. \citenum{mahajan_importance_2021})} & Expt.  \\ \midrule
        $U_1\,$ (eV) & ---             & 1.59          & 1.64        & 6.34           & 6.76 & ---             \\
        $U_2 / V\,$ (eV) & ---             & ---          & 4.62         & ---           & (0.99, 1.10) & ---             \\ \midrule
        $a\,$ (\AA)             & 4.37            & 4.37          & 4.38              & 4.40           & 4.39               & 4.40 \cite{bolzan_powder_1993}            \\
        $c\,$ (\AA)             & 2.83            & 2.86          & 2.88              & 2.94           & 2.92               & 2.88 \cite{bolzan_powder_1993}            \\
        $V_0\,$ (\AA$^3$)       & 54.68           & 54.64         & 55.26             & 57.07          & 56.35              & 55.79 \cite{bolzan_powder_1993}           \\
        $E_\mathrm{g}\,$ (eV)            & 0.00            & 0.00          & 1.01              & 0.02           & 0.32               & \parbox{2.6cm}{\centering 0.26\cite{chevillot_influence_1959}, \\ 1.50 \cite{kumar_morphology_2017}, 2.0\cite{bigiani_dual_2020}} \\ \bottomrule
        \end{tabular}
        \end{table*}
        
\begin{figure}[t]
            \includegraphics[width=3.33in]{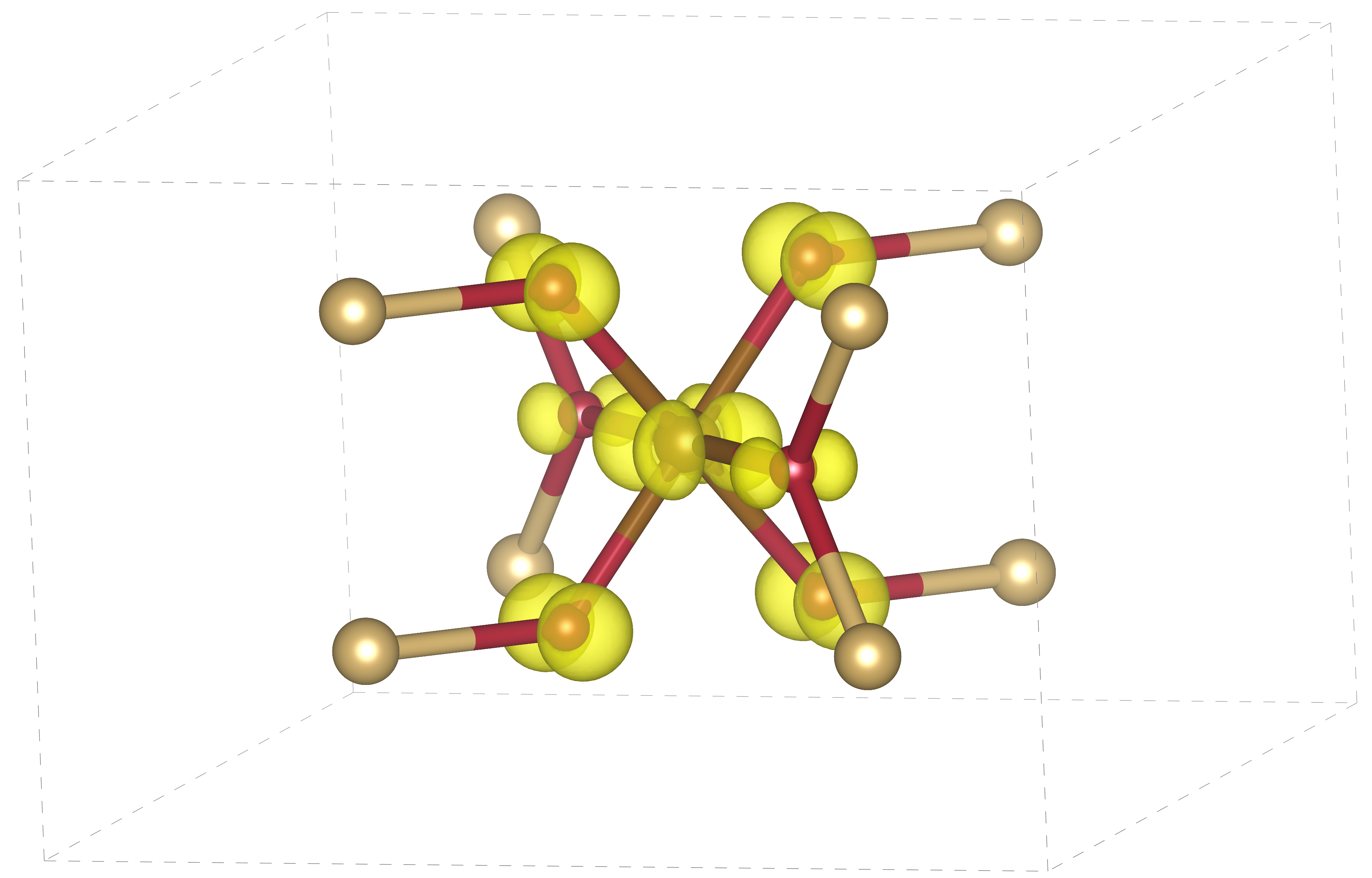}
            \caption{Isosurface plot of the highest fully occupied KS band of A1-AFM \ce{\beta-MnO2} obtained using PBEsol$+U^{\mathrm{Mn}}_{t_{2g}}$, showing $\pi$ interactions between \ce{O}-${p_z}$ and \ce{Mn}-${t_{2g}}$ orbitals. Mn atoms of opposite spin-polarization are colored dark and light brown, O atoms are red. For clarity, the interaction is only shown for one \ce{MnO6} octahedron.}
            \label{fig:vbm-mno2}
        \end{figure}

        The first manifold solely considers the ${t_{2g}}$ orbitals in order to single out the impact of the localized \ce{Mn}-$d$ states. \footnote{Here, we use the $t_{2g}$ irreducible representation in spite of the actual $D_{4h}$ site symmetry of \ce{Mn^{4+}} and therefore neglect the small energetic splitting between the $A_{1g}$, $B_{1g}$ and $B_{2g}$ representations.}
        As shown in Table \ref{tab:mno2-results}, this correction only marginally influences the band gap or the structural properties in comparison to bare PBEsol.
        This observation aligns with the modest value of $U^{\mathrm{Mn}}_{t_{2g}} = 1.59\,$eV, indicative of minimal local curvature due to the ${t_{2g}}$ states.
        A much more significant improvement is achieved by considering the localized \ce{O}-${p_z}$ orbitals (Figure \ref{fig:vbm-mno2}). 
        Correcting these states with a self-consistent Hubbard parameter $U^{\mathrm{O}}_{p_{z}}=4.62\,$eV (in addition to the $U^{\mathrm{Mn}}_{t_{2g}}$ correction) leads to an insulating gap of $1.01\,$eV. 
        Moreover, the lattice parameters approach their experimental values, resulting in an error on the predicted cell volume smaller than 1\%.
        Conversely, a shell-averaged $U$ correction as large as $6.34\,$eV for Mn-$d$ states only results in less accurate cell parameters and a still negligible band gap.
        This situation is only partly mended by the application of an additional inter-site $V$ parameter \cite{mahajan_importance_2021}.

\begin{figure}[t]
            \includegraphics[width=3.33in]{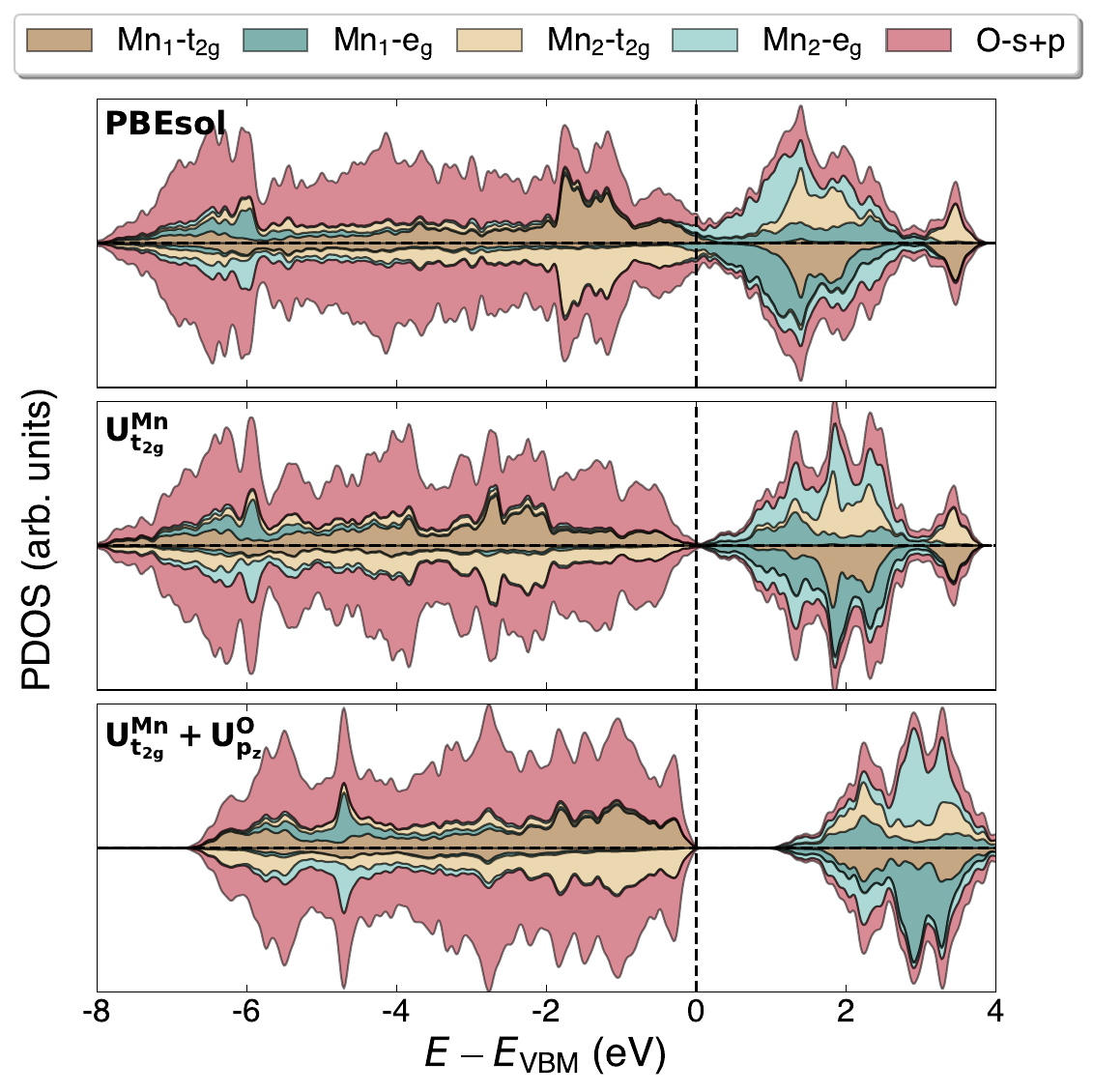}
            \caption{Stacked spin-resolved PDOS of the A1-AFM collinear ordering of \ce{\beta-MnO2} computed with PBEsol and two types of orbital-resolved $U$ corrections. Metallic behavior is observed when using PBEsol and PBEsol$+U^{\mathrm{Mn}}_{t_{2g}}$, whereas an insulating character is obtained with PBEsol$+U^{\mathrm{Mn}}_{t_{2g}}+U^{\mathrm{O}}_{p_z}$. On each panel, the upper and lower subpannels correspond to the spin-up and spin-down channels, respectively.}
            \label{fig:pdos-mno2}
        \end{figure}

        The importance of including \ce{O}-${p_z}$ in the Hubbard manifold is evidenced by the PDOS depicted in Figure~\ref{fig:pdos-mno2}.
        The states near the Fermi level exhibit notable \ce{O}-${p}$ contributions, regardless of the correction applied. These contributions primarily stem from localized ${p_z}$ orbitals, which engage in $\pi$ interactions with the \ce{Mn}-${t_{2g}}$ manifold (Figure \ref{fig:vbm-mno2}).
        The other two $p$ orbitals are subject to $sp^2$ hybridization and contribute to the formation of the lower-lying $\sigma$ MOs in conjunction with \ce{Mn}-${e_g}$ states. 
        This observation underscores that applying shell-averaged $U$ corrections to the entire \ce{O}-$p$ shell would weaken the $\sigma$ bonds and lead to Mn-O underbinding. For the same reason, applying corrections to the \ce{Mn}-${e_g}$ states is neither required nor useful.

        In conclusion, the orbital-resolved approach pinpoints deviations from PWL in the localized \ce{O}-$p_z$ orbitals as the potential root-cause of the absence of a band gap in \ce{\beta-MnO2} when using DFT with semi-local functionals. 
        This outcome underscores the frequently overlooked significance of ligand orbitals in electronic localization and establishes a foundation for future investigations. 
        Again, achieving more comprehensive descriptions of this intricate charge-transfer insulator could entail incorporating the orbital-resolved $U$ approach alongside inter-site $V$ and unlike-spin Hund's $J$ terms.

\section{Outlook: orbital-resolved inter-site Hubbard interactions?}
    \label{sec:final_remarks}
    Hubbard ${U}$ corrections are primarily designed to mitigate the FCE in KS states exhibiting strong atomic-like character, that is, at the on-site level. Orbitals involved in strong inter-atomic hybridization should not be corrected using $U$ terms. This limitation is a consequence of the one-center projectors that are typically used to determine the occupation numbers for the Hubbard $U$ energy functional, which inherently lack the ability to adequately represent localization on hybrid (i.e., molecular) orbitals. 
    To counteract the FCE within covalent environments, the extended DFT+$U \! + \! V$ framework was introduced in Ref.~\citenum{leiria_campo_jr_extended_2010}, where conventional shell-averaged $U$ corrections are augmented with an inter-site term scaled by a Hubbard parameter $V$ in a way that restores ligand hybridization through two-center (dual) occupation numbers.
    However, although the +$V$ correction often improves the predictions of electronic and structural properties compared to plain DFT+$U$, the results presented for \ce{FeS2} and \ce{MnO2} exemplify that it only partially mitigates the impact of shell-averaged $U$ corrections.
    Specifically, inter-site terms cannot rectify a substantial bias in the total energy arising from on-site terms since the values of $U$ (for first-row TM elements) are typically between 4\, and 10\,eV, which is several times greater than typical values of $V$, which amount to $\approx 1\,$eV.
    For instance, the Hubbard $U$ and $V$ parameters of the strong-field molecular complex \ce{[Fe(CNH)_6]^{2+}} from Sec.~\ref{sec:results_and_discussion} are $8.42$ (for Fe-$3d$ states) and $1.12\,$eV (between Fe-$3d$ and C-$2p$ states) for the LS configuration, and $6.57$ and $1.11\,$eV for the HS configuration, respectively. 
    Applying these parameters, one obtains an adiabatic spin energy difference  $\Delta E_{H-L}=-0.42\,$eV that erroneously suggests a HS ground state for \ce{[Fe(CNH)_6]^{2+}} (the CASPT2/CC reference is $+2.82\,$eV, see Figure~\ref{fig:spin-energies}). 

    Given that the majority of spurious contributions to the total energy arise from the shell-averaged form of $U$, the implementation of orbital-resolved DFT+$U \!+ \! V$ is an intriguing prospect, as already hinted in the concluding remarks of Ref.~\citenum{leiria_campo_jr_extended_2010}.
    This extension could exert precise control over electron localization on molecular orbitals, which could be achieved through individual Hubbard $V^{IJ}_{ij}$ parameters. Moreover, such highly tailored but still fully first-principles corrections should allow to eliminate potential conflicts arising from the simultaneous treatment of orbitals by both $U$ and $V$ terms. With regard to the compounds discussed in the present work, the orbital-resolved $U$ corrections of the localized $t_g$/$t_{2g}$ orbitals could be augmented by $V$ terms specific to the interaction of $e_g$ with neighboring ligand orbitals. Note that orbital-resolved inter-site $V$ parameters are readily obtained as a by-product of the calculation of $U$ parameters within the generalized LR-cDFT approach. 
    However, practical calculations with this fully-resolved approach would require the \textit{a priori} assignment of two-center occupation numbers to specific MOs, which involves additional preparatory effort. Desirable will be the conception and implementation of appropriate automation workflows in supporting tools such as AiiDA\cite{huberAiiDAScalableComputational2020a,uhrin_workflows_2021}.
    Alternatively, a simplified approach could involve using the Hubbard $V$ correction in its current (averaged) form while retaining the orbital-dependence of $U$. This would serve as a pragmatic \textit{ad-hoc} solution to the issues of $U$ at the expense of consistency.
    
    Another avenue for extension pertains to Hund's $J$ corrections, whose combination with the orbital-resolved $U$ should be straightforward. Such `unlike-spin' terms play a pivotal role in addressing the fractional spin error that occurs in systems characterized by significant magnetic coupling \cite{himmetoglu_first-principles_2011,orhan_first-principles_2020,linscott_role_2018,mori-sanchez_discontinuous_2009,bajaj_communication_2017,shishkin_dft_2019,shishkin_evaluation_2021}.

\section{Conclusions}
    \label{sec:conclusions}
    In this study, we have introduced an orbital-resolved generalization of the DFT$+U$ functional originally formulated by \citeauthor{dudarev_electron-energy-loss_1998}.
    Our implementation is agnostic about the specific nature of the Hubbard projectors employed and maintains full invariance against basis vector rotations. As a result, calculations involving forces and stresses require no adjustments stemming from this expansion.
    The true potential of the scheme emerges when the required Hubbard parameters are derived from first principles. For this purpose, we have employed an adapted version of the LR-cDFT approach, enabling perturbative calculations with orbital resolution.
    This approach intrinsically incorporates intra-shell screening, which often causes orbital-resolved $U$ parameters to be significantly smaller than their shell-averaged counterparts.
    Provided a proper selection of the target manifold, the orbital resolution therefore enables a more surgical use of DFT+$U$ that avoids overcorrections, as comparative calculations of six \ce{Fe{(II)}} molecular hexacomplexes as well as of the charge-transfer insulators pyrite and pyrolusite underscore. 

    For instance, the orbital-resolved approach effectively addresses the bias in adiabatic spin energies towards HS states observed in shell-averaged DFT$+U$~\cite{mariano_biased_2020}. Particularly noteworthy is its success in accurately predicting spin energies across a diverse spectrum of \ce{Fe{(II)}} hexacomplexes, achieved through selective corrections of the highly localized $t_{2g}$ orbitals.
    Explicit fractional charge calculations on \ce{Fe[CNH]6^{2+}} suggest that these improvements are not coincidental: the orbital-resolved approach reliably counteracts the spurious global curvature of DFT with (semi-)local functionals with respect to the fractional addition or removal of electrons, reducing the FCE by an order of magnitude.
    In contrast, shell-averaged Hubbard $U$ corrections exhibit a mixed performance: while they effectively diminish the FCE in the HS states, their efficacy wanes when addressing the LS states. Here, instead of rectifying the FCE, the convex error characteristic of (semi-)local DFT is converted into a comparably pronounced concave error. The results also suggest that a sole correction of the localized $t_{2g}$ orbitals performs slightly better than a joint correction of $t_{2g}$ and the rather hybridized $e_{g}$ manifold, even when using orbital-resolved $U$ parameters.
    The superiority of the orbital-resolved formulation over the shell-averaged approximation also extends to the charge-transfer insulators  \ce{FeS2} and  \ce{\beta-MnO2}. The correct non-magnetic ground state of the former is only stabilized when orbital-resolved Hubbard parameters are used. 
    The magnitude of the large experimental band gap of the latter can only be achieved by applying a pinpointed Hubbard correction to the frontier \ce{O}-$p_z$ orbitals, whereas standard Hubbard $U$ corrections to the $d$ shell of \ce{Mn} fail to open a significant band gap.
    
    As these examples illustrate, the success of the orbital-resolved formulation predominantly originates from the exclusion of hybridized orbitals from the Hubbard manifold. 
    From a theoretical perspective, the necessity to exclude hybridized orbitals is rooted in the very definition of $U$ as an \textit{on-site} term.
    However, practically implementing this definition poses challenges, particularly given that many DFT$+U$ investigations rely on atomic-like orbitals as Hubbard projectors.
    This approach can lead to occupancy eigenvalues far from 0 or 1, especially in compounds featuring covalent bonds. These fractional values, however, are not related to the electron self-interaction, as they arise from a one-center projector being applied to a two-center phenomenon, namely a molecular orbital.
    In such scenarios, the orbital-resolved approach provides an \textit{ad-hoc} solution that allows to circumvent the potential conflict of on-site corrections with the intricacies of covalently bonded systems.
    
    In a broader context, it is crucial to emphasize that the refined approach presented in this work constitutes just one among various potent (Hubbard) corrections available to DFT. As such, its effectiveness and ability to achieve consistent enhancements relies on a thoughtful and technically well-executed application.
    In practical terms, this is achieved by adopting a more nuanced strategy for the determination of Hubbard manifolds, tailored to the specific Hubbard projectors employed. Such a strategy should consider \textit{localized} ligand orbitals for Hubbard corrections in situations where these act as frontier states.
    In addition, it might be desirable to address the FCE in hybrid (i.e., molecular) orbitals. For this purpose, the use of Wannier functions as Hubbard projectors or the adoption of an extended orbital-resolved DFT$+ U \! + \! V$ approach should be contemplated.
    A pivotal aspect of our future investigations will focus on developing a protocol to automate the choice of Hubbard manifolds based on measurable criteria. This step aims to streamline and standardize the approach's application, ensuring its systematic and effective use across diverse systems and scenarios.

\begin{acknowledgement}
    We thank Louis Ponet for fruitful discussions. EM was funded by the Deutsche Forschungsgemeinschaft (DFG, German Research Foundation) -- Project No. 286518848 -- RTG 2247 $\mathrm{QM}^3$. 
    IT and NM acknowledge support from the Swiss National Science Foundation (SNSF) through its National Centre of Competence in Research (NCCR) MARVEL (Grant No. 205602). 
    NM acknowledges funding from the U Bremen Excellence Chair programme.
    The authors also acknowledge the computing time granted by the Resource Allocation Board and provided on the supercomputers Lise and Emmy at NHR@ZIB and NHR@G\"ottingen as part of the NHR infrastructure (project hbc00053) and the Swiss National Supercomputing Centre (CSCS) under project No.~s1073.
\end{acknowledgement}

\begin{suppinfo}
Shifts of VBM and CBM KS eigenvalues in pyrite due to Hubbard corrections.
\end{suppinfo}
\bibliography{bibliography}
\end{document}